\documentclass[aip,reprint,superscriptaddress,showpacs,showkeys,preprintnumbers]{revtex4}

\usepackage[english]{babel}
\usepackage{graphicx}
\usepackage{hyperref}
\hypersetup{breaklinks=true,colorlinks=true,citecolor=blue,linkcolor=blue,filecolor=blue,urlcolor=blue}
\usepackage{multirow}
\usepackage[version=4]{mhchem}
\usepackage{multirow}
\usepackage{siunitx}

\begin{document}

\preprint{}

\author{Diego Guedes-Sobrinho}
\email{guedes.sobrinho.d@gmail.com}
\affiliation{Instituto Tecnol\'ogico de Aeron\'autica, DCTA, 12228-900, S\~ao Jos\'e dos Campos, Brazil.}

\author{Ivan Guilhon}
\affiliation{Instituto Tecnol\'ogico de Aeron\'autica, DCTA, 12228-900, S\~ao Jos\'e dos Campos, Brazil.}

\author{Marcelo Marques}
\affiliation{Instituto Tecnol\'ogico de Aeron\'autica, DCTA, 12228-900, S\~ao Jos\'e dos Campos, Brazil.}

\author{Lara K. Teles}
\email{lkteles@ita.br}
\affiliation{Instituto Tecnol\'ogico de Aeron\'autica, DCTA, 12228-900, S\~ao Jos\'e dos Campos, Brazil.}

\title{Thermodynamic Stability and Structural Insights for \ce{CH3NH3Pb_{1-x}Si_xI3}, \ce{CH3NH3Pb_{1-x}Ge_xI3}, and \ce{CH3NH3Pb_{1-x}Sn_xI3} Hybrid Perovskite Alloys: A Statistical Approach from First Principles Calculations}




\begin{abstract}

The recent reaching of 20\% of conversion efficiency by solar cells based on 
metal hybrid perovskites (MHP), e.g., the methylammonium (MA) lead iodide, 
\ce{CH3NH3PbI3} (\ce{MAPbI3}), has excited the scientific community devoted to the 
photovoltaics materials. However, the toxicity of \ce{Pb} is a hindrance for 
large scale commercial of MHP and motivates the search of another congener 
eco-friendly metal.
Here, we employed first-principles calculations via density functional theory 
combined with the generalized quasichemical approximation to investigate 
the structural, thermodynamic, and ordering properties of \ce{MAPb_{1-x}Si_{x}I3}, 
\ce{MAPb_{1-x}Ge_{x}I3}, and \ce{MAPb_{1-x}Sn_{x}I3} alloys as pseudo-cubic 
structures.
The inclusion of a smaller second metal, as \ce{Si} and \ce{Ge}, strongly 
affects the structural properties, reducing the cavity volume occupied by 
the organic cation and limitating the free orientation under high temperature 
effects. 
Unstable and metaestable phases are observed at room temperature for 
\ce{MAPb_{1-x}Si_{x}I3}, whereas \ce{MAPb_{1-x}Ge_{x}I3} is energetically 
favored for \ce{Pb}-rich in ordered phases even at very low temperatures.
Conversely, the high miscibility of \ce{Pb} and \ce{Sn} into \ce{MAPb_{1-x}Sn_{x}I3} 
yields an alloy energetically favored as a pseudo-cubic random alloy with 
tunable properties at room temperature.
\end{abstract}



\maketitle

\section{Introduction}

Justified by the imminent scarcity of energy sources based on conventional fossil fuels, 
the recent rise of metal halide perovskites (MHP defined by \ce{ABX3}) as alternative of 
low cost photovoltaic material has excited the community centered around silicon, which 
has been considered the principal element in solar cells\cite{Green_506_2014,Boix_16_2014,Gratzel_838_2014,Sarapov_4558_2016,Ali_718_2018}.
MHP based on the use of lead iodide (\ce{PbI3^{-}}) and methylammonium (\ce{CH3NH3^{+}} 
$=$ \ce{MA^{+}}), i.e., \ce{MAPbI3}\cite{Stranks_391_2015,Manser_12956_2016,Sutherland_295_2016},
reached remarkable 20\%\cite{Cui_036004_2015} of efficiency in lighting conversion 
devices, which has put it as background for improvements of its photovoltaic 
performance\cite{Gottesman_2332_2015,Chen_081902_2016,Kovalsky_3228_2017,Lee_4270_2017,Lee_8693_2017}.
However, a deeper comprehension of the chemical and structural properties correlated 
with the optical efficiency is needed. Additionally, for a large scale commercialization 
of solar cells based on MHP, combining thermodynamic stability and high photovoltaic 
performance is the key point for the viability of those devices\cite{You_75_2015,Raza_20952_2018,Abdelmageed_387_2018}.

Experiments have revealed the \ce{MAPbI3} stability in different structural motifs into
a relative short range of temperatures. For example, below \SI{163}{\kelvin} the orthorhombic
($Amm2$ space group, $a = \SI{8.84}{\AA}$, $b = \SI{12.58}{\AA}$, $c = \SI{8.55}{\AA}$) is found\cite{Baikie_5628_2013},
between \num{163}$-$\SI{328}{\kelvin} the structure becomes tetragonal ($I4/mcm$ space group,
$a = \SI{8.87}{\AA}$, $b = \SI{12.67}{\AA}$), and above \SI{328}{\kelvin}\cite{Sharada_2412_2016} 
\ce{MAPbI3} has been suggested as pseudo-cubic ($P4mm$ space group, $a = \SI{6.31}{\AA}$)\cite{Stoumpos_9019_2013}.
Additionally, the thermodynamic stability of MHP has been investigated aiming their obstacles 
against the inclement weather, such as UV, moisture, heat, and oxygen, which is crucial 
for MHP durability of photovoltaic cells\cite{Qinglong_7617_2015,Joseph_5102_2015,Lee_1_2016,Zhang_036104_2018}.
Experiments of differential thermal analysis has indicated the decomposition of \ce{MAPbI3} 
tetragonal phase in \ce{CH3NH3PbI3}($s$) $\longrightarrow$ \ce{PbI2}($s$) + \ce{CH3NH2}($g$) + \ce{HI}($g$), 
in order that for temperatures from \SI{403}{\kelvin} the perovskite gradually starts to be 
decomposed\cite{Brunetti_31896_2016}.
It is reported through X-ray diffraction that even after the \ce{MAPbI3} systems be 
submitted under temperature of \SI{443}{\kelvin} the sample keeps as \SI{69}{\percent} 
of \ce{MAPbI3} and \SI{31}{\percent} of \ce{PbI2}\cite{Brunetti_31896_2016}.
For \ce{MASnI3}, for instance, X-ray diffraction experiments revealed the presence of 
tetragonal structure at \SI{423}{\kelvin}, and at room temperature by considering an 
\ce{MASn_$x$I3} for $0.9 \leq x \leq 1.4$ as relative quantities between MA:\ce{Sn^{2+}} 
(in 1:$x$) used throughout the synthesis process, the perovskite adopts a pseudo-cubic 
structure for some $x$ values\cite{Dimesso_4132_2017}. 
However, the thermal decomposition starts only at \SI{473}{\kelvin}, which is a higher 
than for \ce{MAPbI3}.

From the last years the mixtures (alloys) \ce{MAPbI3}-based perovskites has provided a new 
perspective to stabilize and tune MHP properties from their composition through several 
different ways, such as: ($i$) changing the \ce{MA^{+}} organic cation by another keeping 
the charge balance\cite{Norman_3151_2015,Weber_15375_2016,Kubicki_14173_2017}; ($ii$) replacing 
\ce{Pb^{2+}} atoms by another cation, e.g., \ce{Sn^{2+}} or \ce{Ge^{2+}}\cite{Ogomi_1004_2014,Hao_8094_2014,Feng_8227_2015,Sampson_3578_2017};
or ($iii$) varying the halogen\cite{Colella_4613_2013,Pathak_8066_2015,Shi_1605005_2017}.
This approach brought a tremendous progress in the development of \ce{MAPbI3}-based for
photovoltaic devices, especially for the \ce{MAPb_{1-x}B_{x}I3} alloys, from which the
toxicity of \ce{Pb} can be suppressed through the use of another congener eco-friendly 
metal (e.g. \ce{B} = \ce{Sn} or \ce{Ge})\cite{Memant_7122_2014,Cortecchia_1044_2016}.
Based on that, those MHP alloys open an enhancement field for the photovoltaic performance 
by chemical control of the thermodynamic stability and optical properties\cite{Stoumpos_9019_2013,Zhang_535_2016,Yang_524_2017,Kopacic_343_2018}.

Even though \ce{MASnI3} has been investigated as an alternative for lead-free perovskite, 
its low power conversion efficiency\cite{Hao_8094_2014} and low oxidation resistence\cite{Ogomi_1004_2014} 
are some motivatory hindrances to workaround from the use of alloys. 
For instance, the \ce{MAPb_xSn_{1-x}I3} stable alloy was recently investigated by 
Hao \textit{et al.}\cite{Hao_8094_2014}, who showed experimentally the control of the band 
gap of the \ce{MAPbI3} (\SI{1.55}{\electronvolt}) for compositions towards \ce{MASnI3} pure 
(\SI{1.30}{\electronvolt}), as the lower band gaps in \SI{1.17}{\electronvolt} and \SI{1.24}{\electronvolt} 
for \ce{MAPb_{0.5}Sn_{0.5}I3} and \ce{MAPb_{0.75}Sn_{0.25}I3}. 
Furthermore, the study revealed that the \ce{MAPb_{0.5}Sn_{0.5}I3} alloy adopts a pseudo-cubic 
structure, while in so far as the content of \ce{Pb} increases the structure adopts a tetragonal 
configuration, i.e., gradually reaching the stable phase of \ce{MAPbI3} at room temperature. 
In others studies focused on the optical and eletrochemical properties\cite{Ogomi_1004_2014,Hu_11436_2017},
it was found an increase for the incident photon wavelength for \ce{MAPb_{0.5}Sn_{0.5}I3}, 
which was red-shifted to \SI{1060}{\nm}, corresponding to the \SI{260}{\nm} displacement 
with respect to the \ce{MAPbI3} pure. 
Beyond that, since a large band gap of \SI{1.90}{\electronvolt} for \ce{MAGeI3} has been 
found,\cite{Stoumpos_6804_2015} the \ce{MAPb_xGe_{1-x}I3} alloy as tetragonal structure 
has also been investigated through theoretical calculations\cite{Sun_14408_2016}. 
The alloys presented narrower band gaps than their pure perovskites counterparts, and even 
though the \ce{MAGe_{0.75}Pb_{0.25}I3} composition has presented the highest theoretical 
efficiency of about \SI{24}{\percent}. However, this study is restricted to few configurations
and a deeper understanding of the structural stability is still needed.

As first attempt to determine the stability of a hibrid perovskite from a specific composition, 
the Goldschmidt`s tolerance factor ($t$) is a geometric parameter initially used to predict 
the ability to form a 3D perovskite\cite{Goldschmidt_477_1926}, which empirically lie into 
$0.80 < t < 1.1$ range\cite{Kieslich_3430_2015,Manser_12956_2016}.
The $t$ is part of an empirical relation given by $\left(R_\text{A} + R_\text{X}\right) = 
t\sqrt{2}\left(R_\text{B} + R_\text{X}\right)$, where $R_\text{A}$ is the effective radii 
of organic cation, $R_\text{B}$ the radii for bivalent metal cation, $R_\text{X}$ for halide 
anion. However, the Goldschmidt`s tolerance factor is limited to predict the perovskite alloys 
formation is limited, since that parameters as the miscibility between the different metals 
involved within the crystal, i.e., concerning the octahedral inner sites occupied by \ce{Pb} 
or a second metal \ce{B}, as well as temperature relative to the thermodynamic favoring 
associated to the alloy stability, are crucial features for the comprehension of their 
electronic and atomic properties in dependence with the composition\cite{Stoumpos_2791_2015}.
Furthermore, the \ce{Pb}/\ce{B} ratio for the metal size creates crystalline distortions 
(combined with the different magnitude for the spin-orbit contributions) which gives important 
insights for electronic characterization of those systems\cite{Isarov_5020_2017,Yang_92_2017,Yaffe_136001_2017}.
As such, a theoretical study for perovskite alloys needs a proper statistical approach 
relative to the configurational sampling constituting the statistical ensemble, which is 
required to calculate the average of thermodynamic and structural properties.

Here, we have performed first-principle calculations based on the Density Functional Theory 
(DFT) to investigate possible environment friendly perovskite \ce{MAPbI3}-based alloys. 
The generalized quasi-chemical approximation (GQCA) was used as statistical method, from 
which thermodynamic properties and averages of the structural parameters can be calculated 
for a wide chemical range at arbitrary temperatures. 
Thus, an improved picture on the perovskite alloys, the \ce{Si}, \ce{Ge}, and \ce{Sn} metals, 
which present different relative atomic size with respect \ce{Pb}, were studied in a pseudo-cubic 
\ce{MAPbI3} structure, considering their local impact on the structure for different direction 
within the crystal is obtained.

\begin{figure}[t!]
\centering
\includegraphics[width=0.99\textwidth]{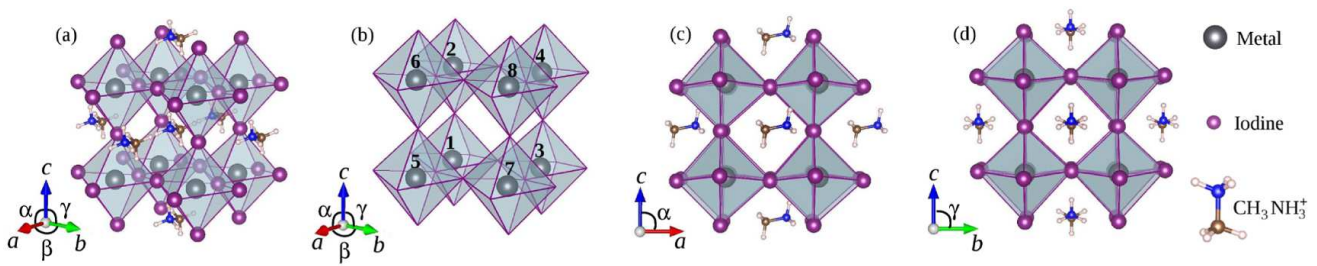}
\caption{(a) Representation of the \ce{MAPb_{1-x}B_{x}I3} cubic supercell for all
the perovskites and alloys based on metals \ce{B} = \ce{Sn}, \ce{Ge}, and \ce{Si}.
(b) \num{8} sites in the \ce{MI3^{-}} octahedrals numbered to replacement of the
metals and formation of the perovskite alloys. (c) Lateral disposition of the organic
cations from the perspective of the $a$ and $c$ directions. (d) Perovskite
with organic cations from the perspective of the $b$ and $c$ directions.
}
\label{fig:super_cell_representation}
\end{figure}

\subsection{Cluster Expansion and Thermodynamic Treatment}

The structural and thermodynamic behaviour of the perovskite alloys were investigated
through a rigorous and systematic statistical description based on the GQCA\cite{Sher_4279_1987}. 
In the GQCA, the alloy (mixture) is described as an ensemble of clusters (herein our supercell),
statistically and energetically independent of the surrounding atomic configuration. 
It has been demonstrated that this model successfully describes the physical properties 
of several 2D and 3D alloys, as well as to 2D sheets\cite{Teles_2475_2000,Schleife_245210_2010,Ghilhon_075435_2015,Freitas_092101_2016}.
Furthermore, the GQCA method also has been employed in the thermodynamic analysis of 
perovskite alloys of \ce{MAPb(I_{1-x}Br_x)_3}\cite{Brivio_1083_2016}, however, the 
method was still not employed for perovsksite alloys from the metal perspective.

Within the GQCA the size and shape of the clusters play an important role, 
wherein the supercell model has two advantages: ($i$) it has a reasonable 
size for taking into account the local correlation; and ($ii$) it has an 
exact counting scheme for the configurational entropy, since no two clusters 
share the same alloying atom.
Based on that, the Figure \ref{fig:super_cell_representation} (a) shows a 
representation of a MHP as a cubic structure (symmetry group $O_h$) with the 
\ce{CH3NH3^{+}} cations balancing the \ce{MI3^{-}} anions charges of the 
octahedrals. 
We used a supercell with $2\times2\times2$ expansion of a cubic perovskite by 
starting from the \ce{MAPbI3} system, from which the alloys are made by replacing 
the \num{8} octahedral central sites by \ce{Sn}, \ce{Ge}, and \ce{Si}, named 
by the letter \ce{B} in the general case, to build the \ce{CH3NH3Pb_{1-x}Sn_xI3}, 
\ce{CH3NH3Pb_{1-x}Ge_xI3}, and \ce{CH3NH3Pb_{1-x}Si_xI3} systems, respectively.

Regarding the \num{8} sites involving the replacement of \num{2} metal species, 
as shown in Figure \ref{fig:super_cell_representation} (b), the total number of 
possible atomic configurations is given by $2^n$, where $n$ is the number of 
sites labeled by \num{12345678}, i.e., resulting on $2^8 = 256$ possible configurations 
for each alloy. However, the \num{256} atomic configurations can be organized in 
$J = 22$ symmetry equivalence classes by considering all the $O_h$ space group 
operations.
The Table \ref{tab:gqca_parameters} describes the \num{22} classes with respect 
the replacement of the octahedral sites, wherein \ce{Pb} atoms are labeled by 
\ce{A} and the \ce{Sn}, \ce{Ge}, and \ce{Si} atoms by \ce{B}.

\begin{table}[t!]
\centering
\begin{tabular}{lcccc|ccccc} \toprule
\multirow{2}{*}{$j$} & \multirow{2}{*}{$n_j$} & Configuration & \multirow{2}{*}{$g_j$} & & & \multirow{2}{*}{$j$} & \multirow{2}{*}{$n_j$}   & Configuration & \multirow{2}{*}{$g_j$} \\ 
       &           & \num{12345678} &     & & &           &         & \num{12345678} &        \\  \hline
\num{1}   &  \num{0}  & AAAAAAAA &  \num{1}   & & & \num{12}  & \num{4} & AAABBBBA & \num{24} \\
\num{2}   &  \num{1}  & AAAAAAAB &  \num{8}   & & & \num{13}  & \num{4} & AABBBBAA & \num{6}  \\
\num{3}   &  \num{2}  & AAAAAABB &  \num{12}  & & & \num{14}  & \num{4} & ABBABAAB & \num{2}  \\
\num{4}   &  \num{2}  & AAAAABBA &  \num{12}  & & & \num{15}  & \num{5} & AAABBBBB & \num{24} \\
\num{5}   &  \num{2}  & AAABBAAA &  \num{4}   & & & \num{16}  & \num{5} & AABBBBAB & \num{24} \\ 
\num{6}   &  \num{3}  & AAAAABBB &  \num{24}  & & & \num{17}  & \num{5} & ABBABABB & \num{8}  \\
\num{7}   &  \num{3}  & AAABABBA &  \num{8}   & & & \num{18}  & \num{6} & AABBBBBB & \num{12} \\
\num{8}   &  \num{3}  & AAABBAAB &  \num{24}  & & & \num{19}  & \num{6} & ABBABBBB & \num{12} \\
\num{9}   &  \num{4}  & AAAABBBB &  \num{6}   & & & \num{20}  & \num{6} & ABBBBBBA & \num{4}  \\
\num{10}  &  \num{4}  & AAABABBB &  \num{8}   & & & \num{21}  & \num{7} & ABBBBBBB & \num{8}  \\
\num{11}  &  \num{4}  & AAABBABB &  \num{24}  & & & \num{22}  & \num{8} & BBBBBBBB & \num{1}  \\ \hline 
\end{tabular}
\caption{\label{tab:gqca_parameters}
The \num{22} different cluster classes of MHP supercells with \num{8} sites in the \ce{MI3^-} 
octahedrals to study perovskite alloys with their $n_j$ \ce{B} atoms (\ce{Sn}, \ce{Ge}, and \ce{Si}).
The sequence \num{12345678} labeling the sites in the cluster can be found in Figure 
\ref{fig:super_cell_representation} (b), where A is \ce{Pb} and \ce{B} are the \ce{Sn}, \ce{Ge}, 
and \ce{Si} atoms to each alloy, where $g_j$ is the degeneracy factor.
}
\end{table}

\begin{figure}[t!]
\centering
\includegraphics[width=0.99\textwidth]{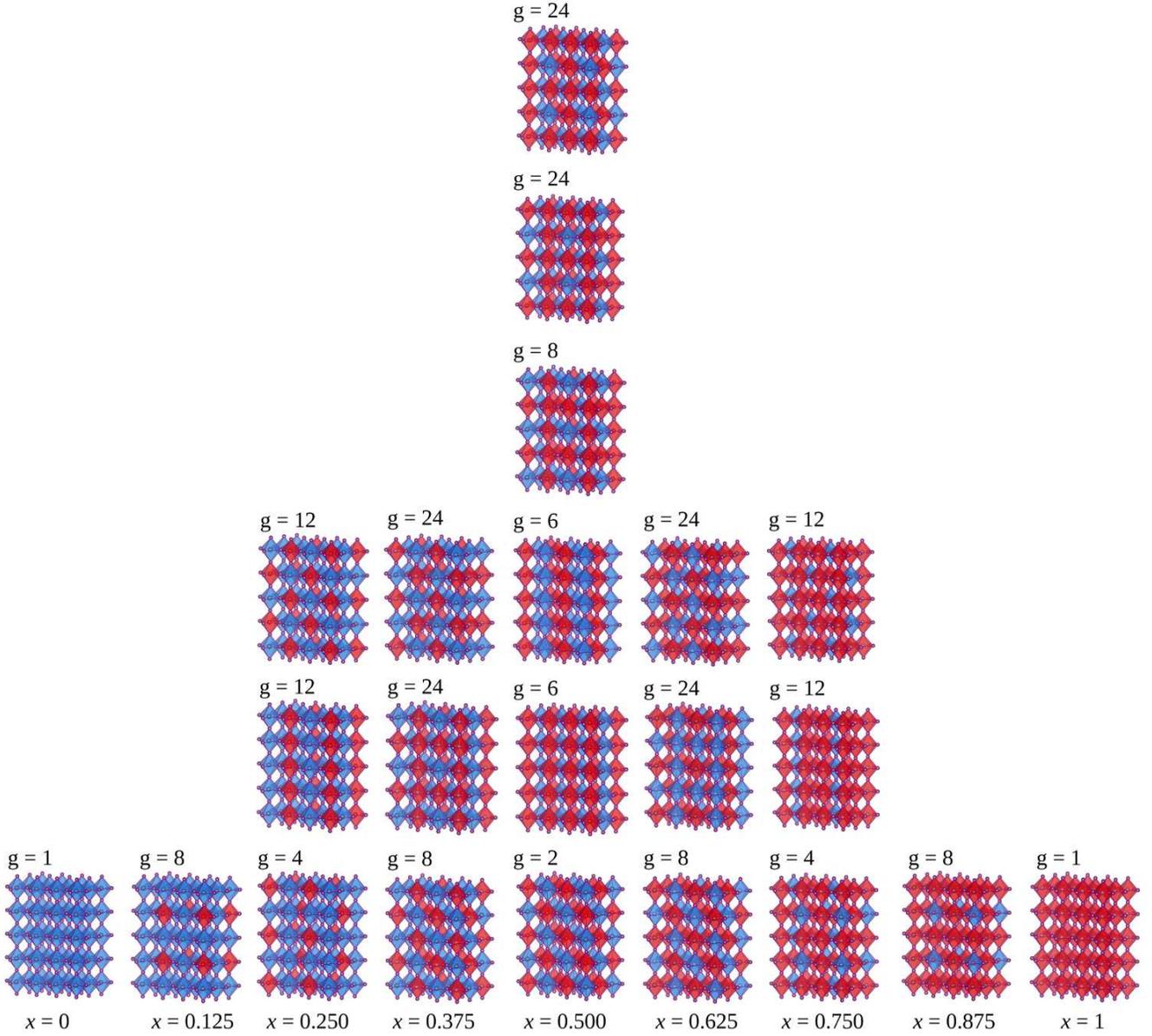}
\caption{\label{fig:all_configurations}
Representations of the \ce{MAPb_{1-x}B_{x}I3} (with \ce{B} = \ce{Si}, \ce{Ge}, and \ce{Si})
isomers for each class $J$ for the $x =$ \num{0}, $0.125$, $0.250$,
$0.375$, $0.500$, $0.625$, $0.750$, $0.875$, \num{1}
compositions. Above each structure, the degeneracy $g$ as used into GQCA method is indicated.}
\end{figure}

The Figure \ref{fig:all_configurations} shows a representation of the relative
positions of the octhedral occupied by \ce{Pb} (blue) and \ce{B} (red) of the
\num{22} classes, as well as their respective compositions $x$ and degeneracies
$g_j$. Thereby, to describe our statistical ensemble for the perovskite alloys, we
considered the set of \num{9} compositions, as $x =$ \num{0}, $\num{0.125}$, $\num{0.250}$,
$\num{0.375}$, $\num{0.500}$, $\num{0.625}$, $\num{0.750}$, $\num{0.875}$, \num{1},
which were defined by the quantities of both metals involved in the alloy formation.
Thus, for a given $N$ as the total number of metals involved (or as the total number
of sites occupied aforementioned), $x = \frac{n_j}{n}$ with $n_j$ as the number of
\ce{Sn}, \ce{Ge}, and \ce{Si} atoms, and $n - n_j$ the number of \ce{Pb} atoms
in the cluster $j$.
Thus, the excess energy of each of those $j$ configurations among the \num{22} 
possibilities with internal mixing energy $\Delta \varepsilon_j$ can be defined by
\begin{equation}
\Delta \varepsilon_j = E_j - (1-x)E_\text{\ce{MAPbI3}} - xE_\text{\ce{MABI3}},
\end{equation}
where, $E_j$, $E_\text{\ce{MAPbI3}}$, and $E_\text{\ce{MABI3}}$ are the total energies
of the cluster configuration $j$, the cluster of \ce{MAPbI3}, and the cluster of 
\ce{MABI3} with \ce{B} = \ce{Pb}, \ce{Sn}, \ce{Ge}, and \ce{Si} pure perovskites. 
As such, the internal energy is calculated by $\Delta U(x,T) = \sum_{j=1}^J x_j(x,T) \Delta \varepsilon_j$, 
where $x_j$ is the probability distribution for the occurence of a cluster with
configuration $j$.
As described elsewhere\cite{Sher_4279_1987,Teles_2475_2000,Ghilhon_075435_2015,Freitas_092101_2016}, 
the occurence probability $x_j$ of equivalence class $j$ can be estimated by the 
constrained minimization of the Helmholtz free energy, i.e., $\Delta F(x,T) = \Delta U(x,T) - T\Delta S(x,T)$,
through the GQCA, by considering the probability normalization $\sum_{j=1}^J x_j(x, T) = 1$ 
and average of composition $x$ as calculated by $\sum_{j=1}^J n_j x_j(x, T) = nx$\cite{Sher_4279_1987,Chen_1995,Teles_2475_2000}.
Thereby, the $x_j(x, T)$ distribution is given by 
\begin{equation}
x_j = \displaystyle \frac{g_j \eta^{n_j} \mathrm{e}^{-\beta \Delta \varepsilon_j}}{\displaystyle \sum_{j=1}^Jg_j \eta^{n_j} \mathrm{e}^{-\beta \Delta \varepsilon_j}},
\end{equation}
where $\beta = (k_\textbf{B}T)^{-1}$, and $\eta$ is an adimensional parameter obtained by
the average composition constrain, and $g_j$ is the degeneracy defined to each $j$ as
described in Table \ref{tab:gqca_parameters}.
The set of probabilities $x_j$ permits to calculate any arbitrary property $p(x,T)$
for the alloy by
\begin{equation}
p(x,T) = \sum_{j=1}^J x_j(x,T) p_j,
\end{equation}
where $p_j$ is the local property of each cluster class $j$.

The mixing entropy in $\Delta S(x,T)$ equation is calculated as
\begin{equation}
\Delta S(x,T) =  -Nk_\textbf{B} \left[ x \text{ln} x + (1-x)\text{ln}(1-x)\right] -Mk_\textbf{B}D_{KL}(x_j || x_j^0), ~~~\textrm{where}~~~ D_{KL}(x_j || x_j^0) = \sum_{j=1}^J x_j \text{ln}\left(\frac{x_j}{x_j^0}\right),
\end{equation}
wherein $k_\textbf{B}$ is the Boltzmann constant and $M$ is the total of clusters.
$D_{KL}(x_j || x_j^0)$ is the Kullback-Leibler (KL) divergence as a relative
entropy measure, which is a function of the $x_j^0$ as the random cluster probability
distribution in an ideal solution defined by $x_j^0 = g_j x^{n_j} (1-x)^{n-nj}$,
and the $x_j$, the probability of an individual cluster belonging to the $j$ class.
Even though previous studies have reported the rotational activity for the 
methylamonium cations under high finite temperature effects\cite{Leguy_1_2015,Chen_355_2015,Franssen_61_2017,Wang_2772_2018},
which correlates with the typical range of synthesis temperature of MHP ($\num{300}-\SI{400}{\kelvin}$)\cite{Colella_4613_2013,Baikie_5628_2013,Leguy_1_2015,DunlapShohl_XXX_2018},
intermittent rotational entropic contributions of the organic cations are not 
considered in our thermodynamic approach. Furhermore, the $\Delta \varepsilon_j$ 
values are predominantly determined by the octahedrals configurations with sites 
occupied by \ce{Pb} or \ce{B} for the clusters $j$, as well as the spin-orbit 
coupling interation used in our calculations which comes only from the metals\cite{Even_205301_2012,Even_2999_2013,Even_10161_2015}.
As such, we set all the cations oriented for the same direction as represented in 
Figure \ref{fig:super_cell_representation} (c) and (d), so that the relative 
directions in $a$, $b$, and $c$ were defined as references for the structural 
analysis.

\section{Results and Discussion}

We discuss the structural parameters, such as lattice parameters (on the orthogonal directions $a$,
$b$, and $c$), local \ce{M-I} distances ($d^{\ce{M-I}}$), angles \ce{I-M-I} and between the 
lattice constants ($\alpha$, $\beta$, and $\gamma$), and the volume ({\AA}$^3$) of the unit cell 
for the \ce{MAPb_{1-x}B_{x}I3} perovskite alloys as a function of the composition and temperature.
By taking the \ce{Pb} atom as reference, the atomic sizes decrease rising in the IV group of the 
periodic table, as \ce{B} = \ce{Sn}, \ce{Ge}, and \ce{Si} which are, respectively, \SI{4.08}, \SI{17.01},
and \SI{32.65}{\percent} smaller with respect to the \ce{Pb} atom.\cite{Kittel_2004}
These differences in the atomic sizes of the metals correlated with the organic cation 
occupying the different cavity sizes made by the octahedrals, taking the relative orientations 
on the $a$, $b$, and $c$ directions (as represented in Figure \ref{fig:super_cell_representation}
(c) and (d)), permit a detailed atomistic comprehension for the pure and alloys perovskites 
in different compositions. Furthermore, a thermodynamic characterization through the mixing 
internal energy ($\Delta U$), mixing entropy ($\Delta S$), excess of free energy ($\Delta F$),
as well as the construction of the $T-x$ phase diagram of the perovskite alloys.

\begin{table}[t!]
\centering
\tabcolsep=0.10cm
\begin{tabular}{lccccccccccccccccccc} \toprule
System   & Space group & & \multicolumn{3}{c}{Lattice ({\AA})}  & & \multicolumn{3}{c}{Angles ($^\circ$)} & & \multicolumn{3}{c}{$d^{\ce{M-I}}$({\AA})} & & \multicolumn{3}{c}{$\phi^{\ce{M-I-M}}$($^\circ$)} & &  Volume ({\AA}$^3$)\\ \hline
 &  SGR & & $a$   & $b$  & $c$       & & $\alpha$   & $\beta$    & $\gamma$ & & $a$   & $b$    & $c$  & &  $a$   & $b$    & $c$  & &  $V$ \\ \hline 
\multirow{2}{*}{\ce{MASiI3}} & \multirow{2}{*}{$P4mm$} & & \multirow{2}{*}{\num{6.18}} & \multirow{2}{*}{\num{6.00}} & \multirow{2}{*}{\num{6.16}} & & \multirow{2}{*}{\num{84}} & \multirow{2}{*}{\num{91}} & \multirow{2}{*}{\num{92}} & & \num{2.61} & \num{2.65} & \num{2.69} & & \multirow{2}{*}{\num{165}} & \multirow{2}{*}{\num{168}} & \multirow{2}{*}{\num{164}}  & &  \multirow{2}{*}{\num{235.29}} \\ 
	& & & & & & & & & & & \num{3.62} & \num{3.38} & \num{3.53} & & & & & & \\ 
\multirow{2}{*}{\ce{MAGeI3}} & \multirow{2}{*}{$P4mm$} & & \multirow{2}{*}{\num{6.20}} & \multirow{2}{*}{\num{6.01}} & \multirow{2}{*}{\num{6.14}} & & \multirow{2}{*}{\num{85}} & \multirow{2}{*}{\num{91}} & \multirow{2}{*}{\num{92}} & &\num{2.70}  &  \num{2.77} & \num{2.80} & & \multirow{2}{*}{\num{166}} & \multirow{2}{*}{\num{167}} & \multirow{2}{*}{\num{163}} & &  \multirow{2}{*}{\num{237.07}}  \\
 & & & & & & &  & & & & \num{3.56} & \num{3.28} & \num{3.42} & & & & & & \\ 
\multirow{2}{*}{\ce{MASnI3}} & \multirow{2}{*}{$P4mm$} & & \multirow{2}{*}{\num{6.30}} & \multirow{2}{*}{\num{6.21}} & \multirow{2}{*}{\num{6.32}} & & \multirow{2}{*}{\num{88}} & \multirow{2}{*}{\num{90}} & \multirow{2}{*}{\num{90}} & &\num{2.91}  & \num{3.12} & \num{3.05} & & \multirow{2}{*}{\num{173}} & \multirow{2}{*}{\num{169}} & \multirow{2}{*}{\num{170}} & &  \multirow{2}{*}{\num{258.25}} \\
 & & & & & & & & & & &  \num{3.43} & \num{3.13} & \num{3.31} & & & & & & \\ 
\multirow{2}{*}{\ce{MAPbI3}} & \multirow{2}{*}{$P4mm$} & & \multirow{2}{*}{\num{6.35}} & \multirow{2}{*}{\num{6.31}} & \multirow{2}{*}{\num{6.40}} & & \multirow{2}{*}{\num{90}} & \multirow{2}{*}{\num{90}} & \multirow{2}{*}{\num{90}} & &\num{3.02} & \num{3.17} & \num{3.18} & & \multirow{2}{*}{\num{173}} & \multirow{2}{*}{\num{167}} & \multirow{2}{*}{\num{167}} & &  \multirow{2}{*}{\num{265.79}}\\
 & & & & & & & & & & &  \num{3.35} & \num{3.17} & \num{3.25} & & & & & & \\ \hline
\end{tabular}
\caption{\label{tab:pure_lattice} Lattice parameters, smallest and largest metal-halide distances 
($d^{\ce{M-I}}$), M-I-M angles ($\phi^{M-I-M}$) with respect to the $a$, $b$, and $c$ directions, 
angles between the lattice constants ($\alpha$, $\beta$, and $\gamma$), space group representation (SGR), 
and volume ($V$) of the unit cell for the \ce{MAPbI3}, \ce{MASnI3}, \ce{MAGeI3}, and \ce{MASiI3} 
perovskites.
}
\end{table}

\subsection{Structural Parameters of the Pure Perovskites and Their Alloys}

\paragraph{Pure perovskites:}
The structural parameters for the \ce{MASiI3}, \ce{MAGeI3}, \ce{MASnI3}, and \ce{MAPbI3} perovskites
are shown in Table \ref{tab:pure_lattice}. All the structures adopt a pseudo-cubic structure
($P4mm$), in order the lattice constant values correlates with the atomic sizes of the metals
into pseudo-cubic structures, i.e., $a$, $b$, and $c$ follow \ce{MASiI3} $<$ \ce{MAGeI3} $<$ \ce{MASnI3}
$<$ \ce{MAPbI3}. Our results are in good agreement with experimental reports, and by comparing 
with studies existing from the literature, for \ce{MAPbI3}\cite{Weller_4180_2015} our calculated 
lattice parameters deviate in $a = \SI{0.47}{\percent}$, $b = \SI{-0.16}{\percent}$, and $c = \SI{1.26}{\percent}$, 
while for \ce{MASnI3}\cite{Hao_8094_2014} in $a = \SI{0.96}{\percent}$, $b = \SI{-0.48}{\percent}$, 
and $c = \SI{0.64}{\percent}$. For \ce{MAGeI3}\cite{Krishnamoorthy_23829_2015}, while our results are 
$a = \SI{1.22}{\percent}$, $b = \SI{-1.88}{\percent}$, and $c = \SI{0.24}{\percent}$ with respect 
to the experimental values, \ce{MASiI3} still need accurated experimental structural parameters to compare.

We found that the smaller atomic size for \ce{Si} and \ce{Ge} when compared with \ce{Pb} contributes
to the decreasing of the lattice constants in up to \SI{4.91}{\percent} (relative to the $b$ direction)
for both \ce{MASiI3} and \ce{MAGeI3} in comparison with \ce{MAPbI3}.
As consequence, their octahedrals are locally more distorted, as can be seen in Table \ref{tab:pure_lattice}
through the differences between the shortest and largest $d^{\ce{M-I}}$ values on all $a$, $b$, and $c$
directions. We found that, in general, throughout the $\ce{Si} < \ce{Ge} < \ce{Sn} < \ce{Pb}$ sequence
for the atomic size the shortest $d^{\ce{M-I}}$ distances increase while the largest $d^{\ce{M-I}}$ 
distances decrease, which is an effect of the competition of the metals into neighbor octahedrals by 
the \ce{I} in the vertice between them.
The angles values between the lattice constants ($\alpha$, $\beta$, and $\gamma$) and for the 
octahedral connections, i.e., $\phi^{M-I-M}$, reveals that for \ce{MASnI3} and \ce{MAPbI3} the local 
distortions are similar, since their atomic sizes for \ce{Sn} and \ce{Pb} are similar. 
However, for \ce{MASiI3} and \ce{MAGeI3} the small metal occuping the octahedral sites promote higher 
deviations $\alpha$, $\beta$, and $\gamma$ angles with respect to the \SI{90}{\degree}, by leading also 
to the decreasing of the $\phi^{M-I-M}$ on all directions also as a local distortion effect on the 
octahedrals.

Our unit cell volume results increasing as $V^{\ce{MASiI3}} < V^{\ce{MAGeI3}} < V^{\ce{MASnI3}} < V^{\ce{MAPbI3}}$ 
in correlation with the metal size, i.e., $\ce{Si} < \ce{Ge} < \ce{Sn} < \ce{Pb}$, suggest the same 
tendency relative to the cavity size where the organic cation is sited. For instance, the relative similarity 
between the \ce{MAPbI3} and \ce{MASnI3} pseudo-cubic structures also can be seen as a similar effect 
of the organic cation orientation on the $a$, $b$, and $c$ lattice directions, yielding a low structural
distortion on the pseudo-cubic motif and a low dependency of the structural parameters on $a$, $b$, 
and $c$ directions with respect to the organic cation orientation. 
Consequently, the largest and shortest $d^{\ce{M-I}}$ values are similar on $b$ for \ce{MAPbI3} (\SI{3.17}{\angstrom}) 
and \ce{MASnI3} (\SI{3.13}{\angstrom}) due to the \ce{CH3} and \ce{NH3} hydrogen, while on $a$ and $c$ 
the \ce{C-N} bond axis its slope effects in the cavity are more pronounced on large and short $d^{\ce{M-I}}$ 
values. Conversely, as an effect of the small metal size and a smaller cavity volume, the stronger distortion 
observed for \ce{MASiI3} and \ce{MAGeI3} by comparing with \ce{MAPbI3} indicates a higher dependency relative 
to the organic cation orientation.

Therefore, we considered the momentary orientation of the organic cation to understand its effects 
on the \ce{MI3^{-}} inorganic octahedra. As such, Figure \ref{fig:super_cell_representation} (c) and (d) 
show that was considered for the \ce{MA^{+}} \ce{C-N} bond axis as momentarily oriented on $a$, giving 
the \ce{C-N} bond axis sloped in the cavity on $b$ and providing lowest energy configuration for the \ce{CH3NH3} 
group as reported by several atomistic simulation studies\cite{Brivio_1083_2016,Tao_1_2017,Jiang_24359_2017}.
Thus, it is reasonable to expect that even though the high temperature effects promote the \ce{MA^{+}}
free reorientation in the cavity for \ce{MAPbI3}, while the reorientation may be slightly limited in the 
\ce{MASiI3} and \ce{MAGeI3} pseudo-cubic structures.

\paragraph{Lattice parameters of the alloyed perovskites:}
\begin{figure}[t!]
\centering
\includegraphics[width=0.99\textwidth]{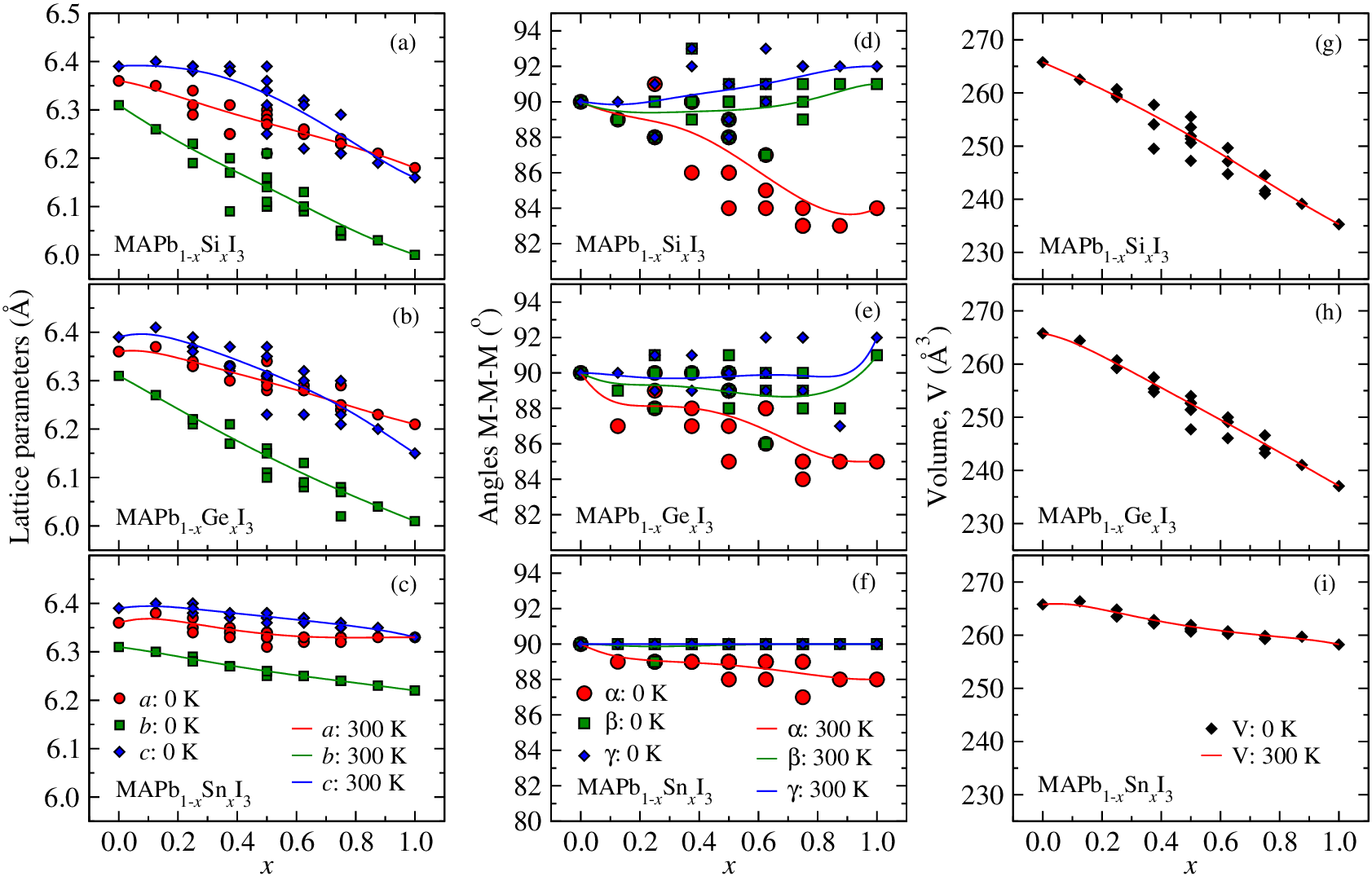}
\caption{Lattice parameters (leftmost) in \SI{}{\angstrom} for the directions $a$, $b$, and $c$, 
angles (middle) between the lattice constants ($\alpha$, $\beta$, and $\gamma$), and volume 
(rightmost) of the unit cell for the \ce{MAPb_{1-x}Si_{x}I3}, \ce{MAPb_{1-x}Ge_{x}I3}, and 
\ce{MAPb_{1-x}Sn_{x}I3} alloys. The symbols filled are the values for the configurations 
$j$ and the solid lines are the average values within the GQCA calculated at \SI{300}{\kelvin}.
}
\label{fig:lattice_parameters}
\end{figure}
The optimization of synthesis process of pure\cite{Colella_4613_2013,Baikie_5628_2013,Sarapov_4558_2016} 
and alloy\cite{Hao_8094_2014,Ogomi_1004_2014,Kopacic_343_2018} MHP at room temperature have widely been 
investigated, especially through self-assembling principles from the chemical precursors for the metal 
halides. As such, our statistic averages were calculated through GQCA at \SI{300}{\kelvin} from the 
weighted contribution of each $j$ configuration, providing the average of the structural parameters 
for the \ce{MAPb_{1-x}B_xI3} alloys as a function of the composition at room temperature.

We calculated the average lattice constants into the supercell on the $a$, $b$, and $c$
directions, as well as the angles between them and the volume for the unit cell for each $j$ cluster 
alloy (Figure \ref{fig:lattice_parameters}). 
Thus, the results connect the values for the \ce{MAPbI3} ($x = 0$) and \ce{MABI3} ($x = 1$), 
\ce{B} = \ce{Si}, \ce{Ge}, and \ce{Sn}. 
We found that the lattice parameters for the \ce{MAPb_{1-x}Si_{x}I3} (panel (a) in 
Figure \ref{fig:lattice_parameters}) alloy follow the Vegard's law\cite{Denton_3161_1991} on the 
$a$ and $b$ directions, i.e., linearly decrease as a chemical specie with smaller atomic size is 
included into the bulk, while for the $c$ direction it is observed a bowing. 
This result is due to the effects of the organic cation orientation 
taken as reference, wherein the \ce{C-N} bond into the small cavity size yields different 
constraints on the lattice on the different directions. For example, on the plane made by $b$ and 
$c$ directions, on which the \ce{C-N} bond of the \ce{CH3NH3^+} is perpendicular, there is a deviation 
of the linearity with respect the composition as an effect of greater permissiveness of lattice 
distance adjustments with respect to the composition. 
Furthermore, as a consequence of the higher contraction of the lattice parameters as the \ce{Si} 
atoms amount increases, we found a crossing over of the lattice parameters on the $a$ and $c$, 
wherein the organic cation orientation yeild lattice distances as $a < c$ and $a > c$ for 
the compositions $x < 0.875$ and $x > 0.875$, respectively.

For the \ce{MAPb_{1-x}Ge_xI3} alloy, the lattice parameter results were similar with the 
\ce{MAPb_{1-x}Si_{x}I3} (panel (b) in Figure \ref{fig:lattice_parameters}). 
We found that the Vegard's law is followed for all the composition range for the $a$ and $b$ 
directions. The crossing over between $a$ and $c$ appears from $x > 0.750$, from which lattice 
parameters are $a > c$. Similarly, this result is also explained for the gradual contraction of 
the lattice parameters due to the small size of the \ce{Ge}, as a consequence of the replacement 
of the \ce{Pb} by \ce{Ge} atoms, by yeilding a decreasing of the cavity size. 
As such, even though the \ce{C-N} atoms of the \ce{MA^{+}} are oriented perpendicular to the 
plane made by $b$ and $c$ orientations, the crossing over between $a$ and $c$ parameters for 
\ce{MAPb_{1-x}Ge_xI3} appears for lower quantities of \ce{Ge} when compared with \ce{MAPb_{1-x}Si_xI3}, 
which is a consequence of larger \ce{Ge} size by comparing with \ce{Si}.

For the \ce{MAPb_{1-x}Sn_xI3} lattice parameters shown into the panel (c) in Figure \ref{fig:lattice_parameters},
since the atomic sizes of the \ce{Pb} and \ce{Sn} atoms are similar there is no crossing over between
$a$ and $c$ parameters, and the Vegard's law is followed in all composition range connecting linearly 
the lattice parameter of the \ce{MAPbI3} and \ce{MASnI3} pure perovskites. 
As such, the linearity connecting the lattice parameters for the \ce{Pb-I-Pb}, \ce{Pb-I-Sn}, or \ce{Sn-I-Sn} 
combinations are independent of the direction, suggesting that the pseudo-cubic structure for 
\ce{MAPb_{1-x}Sn_xI3} alloy is quite resistent with respect to the composition.

\paragraph{Lattice angles and volume of the alloyed perovskites:}
The panels (d), (e), and (f) in Figure \ref{fig:lattice_parameters} show the lattice angles 
($\alpha$, $\beta$, and $\gamma$) for the \ce{MAPb_{1-x}Si_xI3}, \ce{MAPb_{1-x}Ge_xI3}, and 
\ce{MAPb_{1-x}Sn_xI3} alloys as a function of the composition. We found for that the $\alpha$ and $\beta$ 
angles slightly increase between $x = 0$ and $x = 1$ for \ce{MAPb_{1-x}Si_xI3} lying into the interval 
$90^\circ-92^\circ$. Conversely, $\gamma$ decrease sharply with angle from $90^\circ$ up to $84^\circ$, 
which is explained by the strong distortion on the pseudo-cubic structure due to the gradual replacement 
of \ce{Pb} by \ce{Si} atoms. Additionally, the volume of the unit cell for the \ce{MAPbI3} and \ce{MASiI3} 
pure perovskites are linearlly connected as function of the composition, with values lying between 
\SI{265.79}{\angstrom^3} and \SI{235.29}{\angstrom^3}, which describes the constraction of 
the alloy by correlating with the Vegard's law.

For \ce{MAPb_{1-x}Ge_xI3} alloys, we found that $\alpha$ and $\beta$ are close to $90^\circ$ between 
$x = 0$ and $0.875$, while $\gamma$ decrease sharply similarly with respect to the \ce{MAPb_{1-x}Si_xI3}, 
that is between $90^\circ$ up to $85^\circ$. This result shows the effects of the metals size differences, 
as well as the linear contraction for the volume of the unit cell between \ce{MAPbI3} and \ce{MAGeI3}.
This behaviour is also indicated for the $\alpha$ and $\beta$ kept in $90^\circ$ for the \ce{MAPb_{1-x}Sn_xI3}
due to the similar size by comparing \ce{Pb} and \ce{Sn}, while $\gamma$ lie into a short interval between 
$89^\circ-90^\circ$.

\begin{figure*}[t!]
\centering
\includegraphics[width=0.99\textwidth]{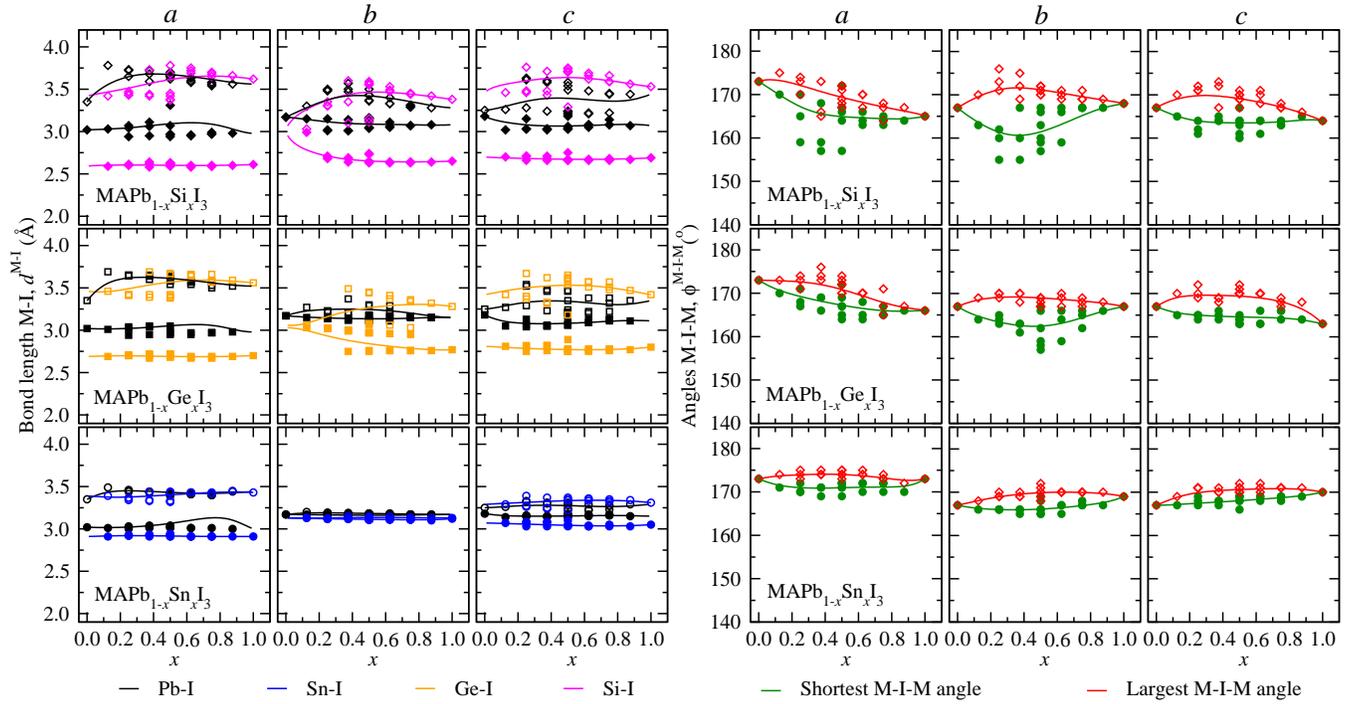}
\caption{Shortest (filled symbols) and largest (empty symbols) \ce{M-I} distances by \ce{M-I} pair, i.e., $d^{\ce{M-I}}$ in \SI{}{\angstrom} 
(\ce{M} = \ce{Si} \ce{Ge}, \ce{Sn}, and \ce{Pb}), and \ce{M-I-M} angles each cluster $j$, as
$\phi^{\ce{M-I-M}}$ in \SI{}{\degree} (degrees), for the \ce{MAPb_{1-x}Si_{x}I3}, \ce{MAPb_{1-x}Ge_{x}I3}, 
and \ce{MAPb_{1-x}Sn_{x}I3} systems with respect to the directions $a$, $b$, and $c$, as a function of the 
alloy composition.
The solid lines are the average values calculated within the GQCA calculated at \SI{300}{\kelvin}.
}
\label{fig:MI_distance_angle}
\end{figure*}

\paragraph{\ce{M-I} distances and \ce{M-I-M} angles as local structural parameters:}
To quantify the structural properties locally for the \ce{MI3^{-}} octahedrals with respect the
alloy compositions, we calculated their shortest and largest \ce{M-I} distances ($d^{\ce{M-I}}$)
and \ce{M-I-M} angles ($\phi^{\ce{M-I-M}}$) on the ($a$), ($b$), and ($c$) directions 
(Figure \ref{fig:MI_distance_angle}).
Once a supercell model was used in our calculations, the $\phi^{\ce{M-I-M}}$ lie into different
values between the shortest and largest \ce{M-I-M} angles. Thus, the plotted $\phi^{\ce{M-I-M}}$ 
values permit to describe the maximum amplitude of the local distortions relative to the compositions 
between $x = 0$ and $x = 1$.
The Figure \ref{fig:MI_distance_angle} shows the $d^{\ce{M-I}}$ and $\phi^{\ce{M-I-M}}$ values
for the \ce{MAPb_{1-x}Si_{x}I3}, \ce{MAPb_{1-x}Ge_{x}I3}, and \ce{MAPb_{1-x}Sn_{x}I3} systems as
a function of the composition. These averages calculated correspond to the equilibrium point 
relative to the equatorial anharmonic octahedral motion of the iodine atom in \ce{M-I-M}\cite{Carignano_20729_2017}.

The shortest $d^{\ce{M-I}}$ values (Figure \ref{fig:MI_distance_angle} leftmost) in the alloys are 
determined by the \ce{Si-I}, \ce{Ge-I}, and \ce{Sn-I} distances, which is an effect of the metal 
size differences with respect to the size of the \ce{Pb}. One observes that for the $a$ and $b$ 
directions that \ce{Pb}-rich compositions the largest $d^{\ce{Pb-I}}$ values are higher than 
$d^{\ce{M-I}}$ values, wherein for few quantities of \ce{B} the shortening of the \ce{B-I} distance 
in an particular octahedral results in an elongation for the \ce{Pb-I} distance relative to the 
neighbor octahedral. 
Thanks to these differences for the metal sizes into the clusters $j$, one observes an increasing 
of the amplitude for the shortest and largest $d^{\ce{M-I}}$ splitted from $\ce{MAPb_{1-x}Sn_{x}I3} 
\rightarrow \ce{MAPb_{1-x}Ge_{x}I3} \rightarrow \ce{MAPb_{1-x}Si_{x}I3}$. This behaviour is explained 
by the local distortions on the octahedrals as the metal size differences are pronounced, also 
as an evidence of the organic cation influence on the organic lattice since the volume of the 
cavity decreases from $x = 0$ to $x = 1$.
Furthermore, except for the $d^{\ce{M-I}}$ values for \ce{Pb}-rich composition on $a$ direction, 
our results show that the shortest and largest $d^{\ce{Pb-I}}$ values tends to keep as the those 
ones in the \ce{MAPbI3} pure perovskite, while the $d^{\ce{B-I}}$ values converge to the \ce{MABI3} 
pure values even for few quantities of \ce{B}.

The $\phi^{\ce{M-I-M}}$ values for each cluster $j$ on all directions (Figure \ref{fig:MI_distance_angle} 
rightmost) highlight distortions into the pseudo-cubic alloys, herein stronger as the difference 
between the metals involved increases. For instance, for \ce{MAPb_{1-x}Si_{x}I3} the average 
$\phi^{\ce{M-I-M}}$ values lie between \SI{165}{\degree}-\SI{175}{\degree}, \SI{160}{\degree}-\SI{175}{\degree}, 
and \SI{165}{\degree}-\SI{170}{\degree} on the $a$, $b$, and $c$ directions, respectively. 
One observes the effects of the strong local distortions induced by the presence of metals 
so different in size, e.g., \ce{Pb} and \ce{Si}, so that there is no linear correlation between 
the \ce{MAPbI3} and \ce{MASiI3} in the alloy formation. The \ce{MAPb_{1-x}Ge_{x}I3} alloy presents 
into softer distortion when compared with \ce{MAPb_{1-x}Si_xI3}, as observed by the 
$\phi^{\ce{M-I-M}}$ values into \SI{165}{\degree}-\SI{175}{\degree}, \SI{165}{\degree}-\SI{170}{\degree}, 
and \SI{165}{\degree}-\SI{170}{\degree} intervals on the, respectively, $a$, $b$, and $c$ directions.
Moving to \ce{MAPb_{1-x}Sn_{x}I3}, the $\phi^{\ce{M-I-M}}$ values are similar from both \ce{MAPbI3} 
and \ce{MASnI3} pure perovskites, in order that small deviations appear between \SI{170}{\degree}-\SI{175}{\degree}
on the $a$ direction and between \SI{165}{\degree}-\SI{170}{\degree} on the both $b$ and $c$ directions.

With the results above discussed, we note the important role of the atomic size difference
between the metals involved in the perovskite alloy formation. For \ce{MAPb_{1-x}Sn_{x}I3},
as a case of similar size for the metals, the small local distortions into the octahedral
and the linearity correlation between the \ce{MAPbI3} and \ce{MASnI3} pure perovskites show
a preference in preserving the pseudo-cubic structure similar to the pure perovskites in the 
whole range of compositions.
Conversely, the \ce{MAPb_{1-x}Si_{x}I3} and \ce{MAPb_{1-x}Ge_xI3} alloys are examples of 
large difference between the atomic size of the metals, we found that the composition is 
an additional variable with respect to the temperature to promotes strong distortions into 
the phase, reinforcing the necessity of a proper statistical analysis to correlates the 
thermodynamic stability with the structural motifs for the alloy.

\subsection{Thermodynamic Parameters and Ordering Preference}

To predict the most favorable local arrangement of metal in the octahedral inner sites, i.e.,
the \ce{PbI3^{-}} and \ce{BI3^{-}} relative configuration, the alloy excess energies ($\Delta \varepsilon_j$)
were calculated in order to determine the composition-dependent cluster probabilities ($x_j$).
Consequently, by knowing $x_j$ as dependent of $\Delta \varepsilon_j$ and the degeneracies $g_j$
for each $j$-configuration, we calculate the mixing free energy $\Delta F\left(x, T\right)$
from the contributions of the interplay between the configurational entropy $\Delta S\left(x, T\right)$
and the internal energy $\Delta U\left(x, T\right)$  through the GQCA. As such, below we
provide a thermodynamic discussion to enlighten the preferential ordering correlated and
the macroscopic stability of the \ce{MAPb_{1-x}B_{x}I3} perovskite alloys.

\begin{figure}[t!]
\centering
\includegraphics[width=0.8\textwidth]{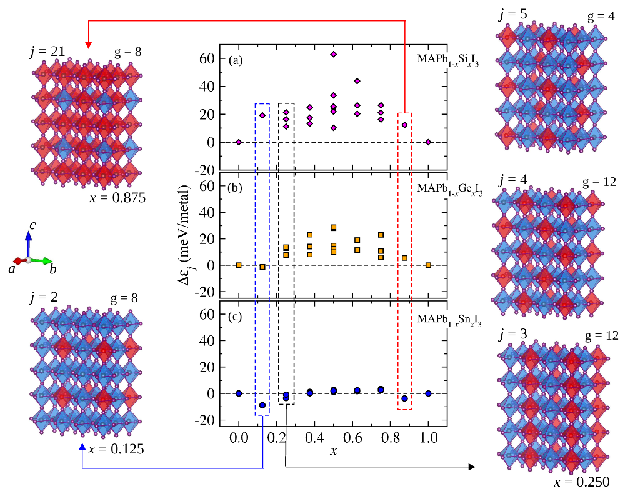}
\caption{Excess energy (midle) in \SI{}{\electronvolt/cluster} for the each configuration $j$ for the
\ce{MAPb_{1-x}Si_{x}I3}, \ce{MAPb_{1-x}Ge_{x}I3}, and \ce{MAPb_{1-x}Sn_{x}I3} perovskite alloys.
The \ce{MA} were omitted for the representations of \ce{PbI3^{-}} (blue octahedrals) and \ce{BI3^{-}}
(red octahedrals).
The blue (leftmost) and red (rightmost) dashed boxes guide to the representation of the ordering 
for $x = 0.125$ ($j = 2$) and $x = 0.875$ ($j = 21$) with degeneracy $g = 8$. Rightmost are the 
configurations $j  = 3$, $4$, and $5$ (black dashed box) for $x = 0.250$ with degeneracies in 
$g = 12$, $12$, and $4$, respectively.
}
\label{fig:excess_energy}
\end{figure}

\paragraph{Alloy excess energies:}
The Figure \ref{fig:excess_energy} provides a plot of the $\Delta \varepsilon_j$ values for the
\num{22} considered cluster configurations as a function of the \ce{B}, i.e., the metals \ce{Si},
\ce{Ge}, and \ce{Sn}, as well as the arrangement representations (omitting the \ce{MA^{+}} cations) of few
configurations and their $g_j$ values for some compositions $x$.
For the \ce{MAPb_{1-x}Ge_{x}I3} and \ce{MAPb_{1-x}Sn_{x}I3} alloys, panels (b) and (c), respectively, 
the most energetically favorable configuration is for $x = 0.125$, as represented by the arrangement 
correspondent to the $\Delta \varepsilon_j$ values indicated by the blue dashed box in Figure \ref{fig:excess_energy}.

The panel (a) shows all the positive $\Delta \varepsilon_j$ values for the \ce{MAPb_{1-x}Si_{x}I3} alloy, 
which means that at $T = \SI{0}{\kelvin}$ there is a high stability of the \ce{MAPbI3} and \ce{MASiI3} 
pure perovskites in detriment of the alloy. 
We found that all the pseudo-cubic configurations strongly distorted between $0 < x < 1$ lie into 
$\Delta \varepsilon_j$ values between \SI{10}{} and \SI{63}{m\electronvolt/metal}, which is an 
evidence of the high strain yielded by the difference of the atomic size between \ce{Pb} and \ce{Si}.
By comparing with the \ce{Ge} alloy, in panel (b), an energetically favored cluster with 
$\Delta \varepsilon_j = \SI{-1.28}{m\electronvolt/metal}$ is observed at $x = 0.125$, which 
correlates with a tendency to form a long-range ordered alloy depending on the temperature.
However, all the distorted pseudo-cubic configurations for $0.125 < x < 1$ present $\Delta \varepsilon_j$ 
values between \num{7} and \SI{30}{m\electronvolt/metal} for \ce{MAPb_{1-x}Ge_{x}I3}.
This result suggests that for an \ce{MAPb_{1-x}B_xI3} (with \ce{B} = \ce{Si}, \ce{Ge}, or \ce{Sn}) 
perovskite alloy energetivally favorable two stability parameters are correlated: ($i$) the proportion 
(composition for the alloy) between the metals occupying the octahedral sites; and the ($ii$) magnitude 
of the atomic size difference between the metals involved.

As a consequence of small difference between the atomic size for \ce{Pb} and \ce{Sn} in the 
\ce{MAPb_{1-x}Sn_{x}I3}, the $\Delta \varepsilon_j$ values lie in an interval of energies 
between \num{-9} and \SI{4}{m\electronvolt/metal}. 
Thus, several configurations can be easily favorable when the entropy effects be considered. 
Therefore, as previously discussed for the structural parameters, such as the lattice parameters, 
$d^{\ce{M-I}}$, and $\alpha^{\ce{M-I-M}}$ as a function of the $a$, $b$, $c$ directions, this 
results suggest that the replacement of \ce{Pb} by \ce{Sn} yields only slight changing in the 
\ce{MAPb_{1-x}Sn_{x}I3} structure.
Among all the configurations between $x = 0$ and $x = 1$ for the short range of $\Delta \varepsilon_j$
values for the \ce{MAPb_{1-x}Sn_{x}I3} alloy, additionally to the $x = 0.125$ (\SI{-8.82}{m\electronvolt/metal})
and $x = 0.875$ (\SI{-3.77}{m\electronvolt/metal}) compositions showed in Figure \ref{fig:excess_energy},
the three possible configurations at $x = 0.250$ are represented by $j = 3$, $4$, and $5$,
which present $\Delta \varepsilon_j$ in \SI{-3.43}{}, \SI{-0.88}{}, and \SI{-0.82}{m\electronvolt/metal}
respectively. We observe that the ordering $j = 3$ as represented in Figure \ref{fig:excess_energy} 
is the most favored, in which the stability is reached by the stacking of the intercalated \ce{PbI3^{-}} 
and \ce{SnI3^{-}} octahedral rows. 
\begin{figure*}[t!]
\centering
\includegraphics[width=0.99\textwidth]{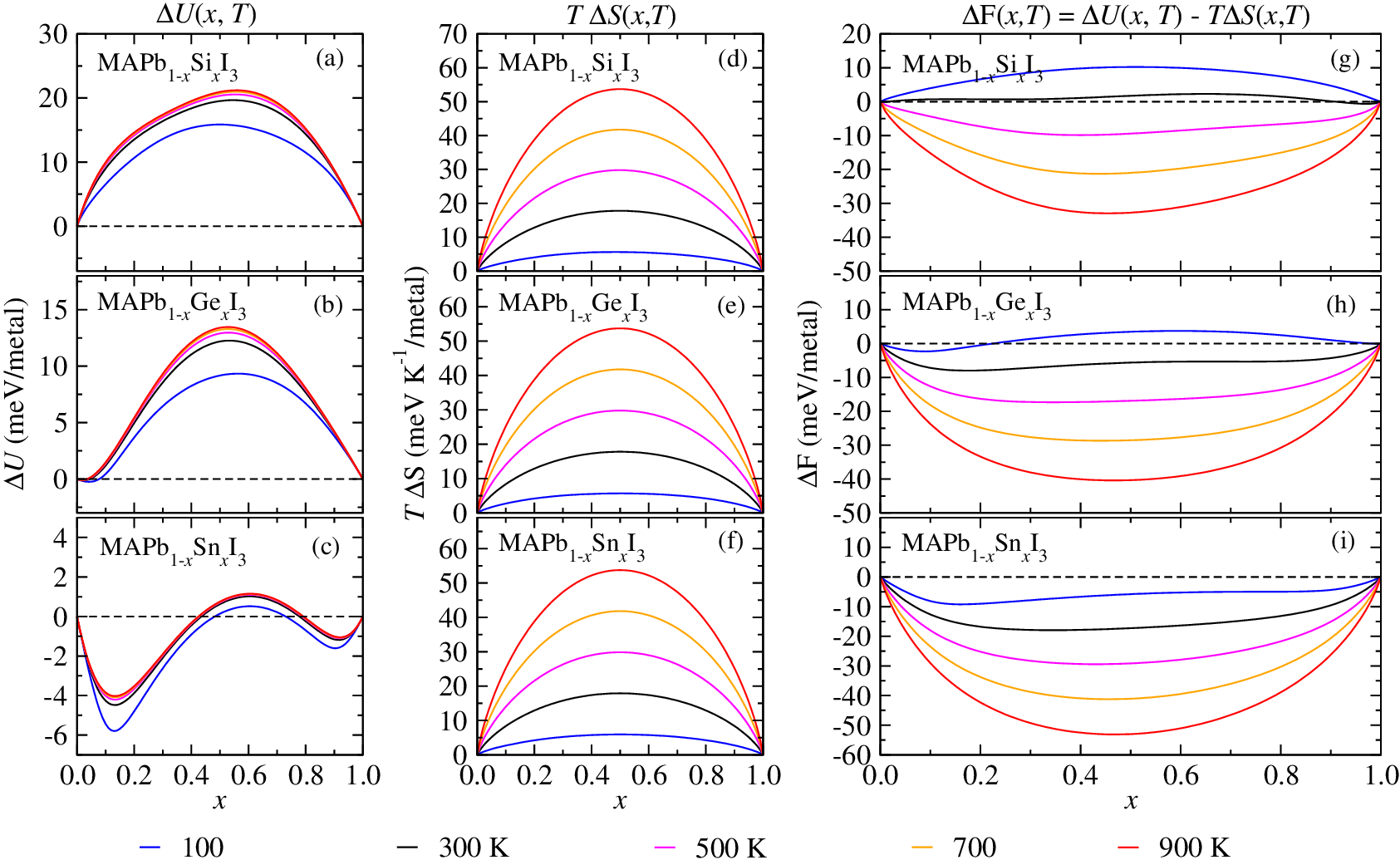}
\caption{Thermodynamic parameters as a function of the alloy composition and temperature for
\ce{MAPb_{1-x}Si_{x}I3}, \ce{MAPb_{1-x}Ge_{x}I3}, and \ce{MAPb_{1-x}Sn_{x}I3} calculated within the
GQCA at \SI{100}, \SI{300},  \SI{500},  \SI{700}, and  \SI{900}{\kelvin}.
Panels (a), (b), and (c) are the averages of the internal energies in \SI{}{m\electronvolt/metal}
($\Delta U$); panels (d), (e), and (f) are the averages of the entropy contribution as a function of the
temperature in \SI{}{m\electronvolt \kelvin^{-1}/metal} ($T\Delta S$); and panels (g), (h), and (i)
are the Helmholtz free energy in \SI{}{m\electronvolt/metal} ($\Delta F$).
}
\label{fig:excess_free}
\end{figure*}

\paragraph{Perovskite alloys free energies and ordering:}
Here, we discuss the statistical contributions of the $\Delta \varepsilon_j$
values for the thermodynamic properties for the alloys under the temperature effects through the GQCA 
method. The variation in the energy of mixing ($\Delta U$) and entropy of mixing ($\Delta S$) used to 
calculate the Helmholtz free energies ($\Delta F$ in \SI{}{m\electronvolt/metal}) for the \ce{MAPb_{1-x}Si_{x}I3},
\ce{MAPb_{1-x}Ge_{x}I3}, and \ce{MAPb_{1-x}Sn_{x}I3} alloys within the GQCA are shown in Figure \ref{fig:excess_free}.
In order to verify the entropy effects for the stabilities of the alloys, we analysed these parameters as
a function of low and high temperatures, e.g., \num{100}, \num{300}, \num{500}, \num{700}, and \SI{900}{\kelvin}.

One observes by the $\Delta S$ symmetrical curves with temperature around $x = 0.500$,
panels (d), (e), and (f), indicating that all the alloys follow an ideal entropy expression at high 
temperatures, i.e., $-Nk_\textbf{B}\left[x \text{ln} x + \left(1 - x\right) \text{ln} \left(1 - x\right)\right]$.
The $\Delta U$ curves for \ce{MAPb_{1-x}Si_{x}I3} $-$ panel (a) $-$ present a positive parabolic 
behaviour due to the higher stability of the \ce{MAPbI3} and \ce{MASiI3} pure perovskites in comparison 
with the alloy. 
Thus, the profile of the $\Delta U$ and $\Delta S$ curves indicates that the alloy can be stabilized 
by entropic contributions, consequently by increasing the magnitude of disorder through the insertion 
of \ce{Si} atoms, which promotes the contribution of several $j$ configurations.
The panel (g) shows a behaviour slightly asymmetric for the $\Delta F$ curve around $x = 0.500$, 
so that for $T < \SI{300}{\kelvin}$ we found $\Delta F > 0$ as an evidence of the instability of the 
alloy at low temperatures.
However, for $T > \SI{300}{\kelvin}$ one observes that $\Delta F < 0$ and the alloy starts to be 
stable, and for temperatures between $\SI{300}{\kelvin} < T < \SI{500}{\kelvin}$ there are points 
throughout $\Delta F$ with same tangent, indicating the existence of a miscibility gap for an extensive 
range of temperatures.

For the \ce{MAPb_{1-x}Ge_{x}I3} alloy, we found that the $\Delta F$ $-$ panel (h) $-$ presents
points with same tangent for $\SI{100}{\kelvin} < T < \SI{500}{\kelvin}$, which is a range of
lower temperatures for the miscibility gap than for \ce{MAPb_{1-x}Si_{x}I3}. 
The $\Delta F$ reaches symmetrical curves for temperatures higher than \SI{500}{\kelvin}, which the 
entropy effects start to be dominant over the small negative $\Delta U$ values, panel (b), for few 
\ce{Ge} quantities. Conversely, with the increasing of the temperature, the disordering is reached 
with the weighted contributions of all \ce{PbI3^{-}} and \ce{GeI3^{-}} octahedrals configurations, 
from which the random configurations for $x = 0.500$ compositions are the most favorable.

The $\Delta U$ curves profile for the \ce{MAPb_{1-x}Sn_{x}I3} alloy $-$ panel (c) $-$ show
the effect of the favorable ordering for compositions with excess of both \ce{Pb} and \ce{Sn}
metals, as $x = 0.125$, and $0.875$.
Firstly, this yields two regions for $\Delta U < 0$ relative to the orderings as represented
in Figure \ref{fig:excess_energy}, so that the alloy stabilizes when the \ce{SnI3^{-}}
individual octahedrals are completely involved by \ce{PbI3^{-}} octahedrals, as well as for 
the opposite configuration, i.e., \ce{PbI3^{-}} individual octahedrals completely involved 
by \ce{SnI3^{-}}. Secondly, the short range of excess energies for \ce{MAPb_{1-x}Sn_{x}I3} 
yields a short interval of $\Delta U$ variation as a function of the composition and temperature.
Thus, for temperatures higher than \SI{100}{\kelvin} the entropy effects are dominant, so 
that the shape of the $\Delta F$ curve becomes more symmetric in order that the contribution 
of all configurations increases with the temperature, consequently, increasing the disordering 
of \ce{PbI3^{-}} and \ce{SnI3^{-}} positions in the alloy.
As such, it is expected to observe a miscibility gap in \ce{MAPb_{1-x}Sn_{x}I3} alloy with
pseudo-cubic structure only for very low temperatures, since there is no effective variation
of the structural environment when the \ce{Pb} in the octahedral sites are replaced by \ce{Sn},
which is a result of the almost similar atomic size between both metals.

\begin{figure}[t!]
\centering
\includegraphics[width=0.99\textwidth]{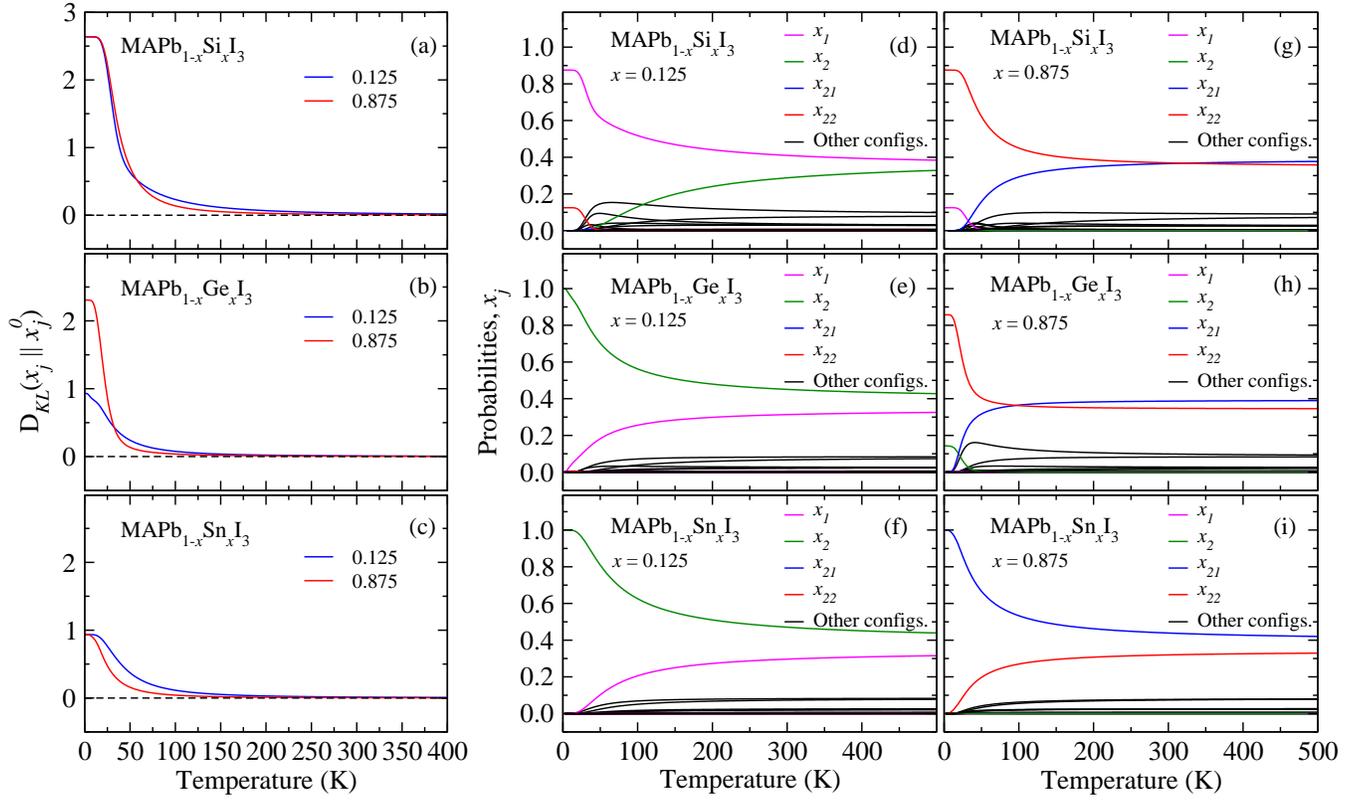}
\caption{
(Leftmost) Kullback-Leibler divergence -- $D_{KL}$($x_j || x_j^0$) -- for all the alloys between
the ideal solid solution and GQCA probability distributions and probabilities $x_j$ (rightmost)
for the ordering $j = 1$, $2$, $21$, and $22$ as a function of temperature and compositions at
$x = 0.125$ and \num{0.875}.
}
\label{fig:dkl_probabilities}
\end{figure}
To investigate the similarity between the GQCA probability $x_j (x,T)$ and a random alloy, relative
to the random contribution of a particular $j$ configuration in a range of temperatures, we present
the KL divergence, namely, $D_{KL}$($x_j || x_j^0$), Figure \ref{fig:dkl_probabilities}.
The maximum divergence at low temperatures means that in a particular composition the configuration
$j$ relative to the $x_j$ dominates over the others, and in so far as the temperature increases the
divergence goes to zero, i.e., the system starts to behave as a random alloy.
In Figure \ref{fig:dkl_probabilities} (a), (b), and (c) are plotted the $D_{KL}$($x_j || x_j^0$)
for \ce{MAPb_{1-x}Si_{x}I3}, \ce{MAPb_{1-x}Ge_{x}I3}, and \ce{MAPb_{1-x}Sn_{x}I3} for the compositions
at $x = 0.125$ and $0.875$ as a function of the temperature.
For \ce{MAPb_{1-x}Si_{x}I3} at very low temperatures one observes a tendency for phase segregation
with the formation of \ce{MAPbI3} and \ce{MASiI3} pure perovskites, as observed through $x_j$
plots in Figure \ref{fig:dkl_probabilities} (d) and (g), wherein there is a predominancy of $x_1$
and $x_{22}$ configurations at compositions $x = 0.125$ and $0.875$, respectively.
For \ce{MAPb_{1-x}Ge_{x}I3}, panel (b), clearly it is seen that at $x = 0.125$ the divergence is
smaller than for $x = 0.875$ at very low temperatures, so that the ordering given by the $x_2 = 1$
configuration dominates at $x = 0.125$ composition for the alloying at \ce{Pb}-rich compositions 
(panel (e)), as well as yields a small contribution at $x = 0.875$ together with the dominant 
$x_{22}$ configuration (panel (h)).
Conversely, for \ce{MAPb_{1-x}Sn_{x}I3} at $x = 0.125$ and $x = 0.875$ and at low temperature,
panel (c), one observes the high miscibility between \ce{Pb} and \ce{Sn} as an effect of the
similar metal size.
Thereby, the $x_j$ plots, panels (f) and (i), show the predominancy of the $x_2 = 1$ ($x_{21} = 1$) 
at $x = 0.125$ ($x = 0.875$) at \SI{0}{\kelvin}, demonstrating the tendency of the system in 
organizing itself in energetically favored alloyed configurations even at very small temperatures. 
The observed ordering of atomic distribution, however, does not persist for temperatures above 
\SI{150}{\kelvin}.

\paragraph{Phase diagram of the alloys:}

\begin{figure}[t!]
\centering
\includegraphics[width=0.99\textwidth]{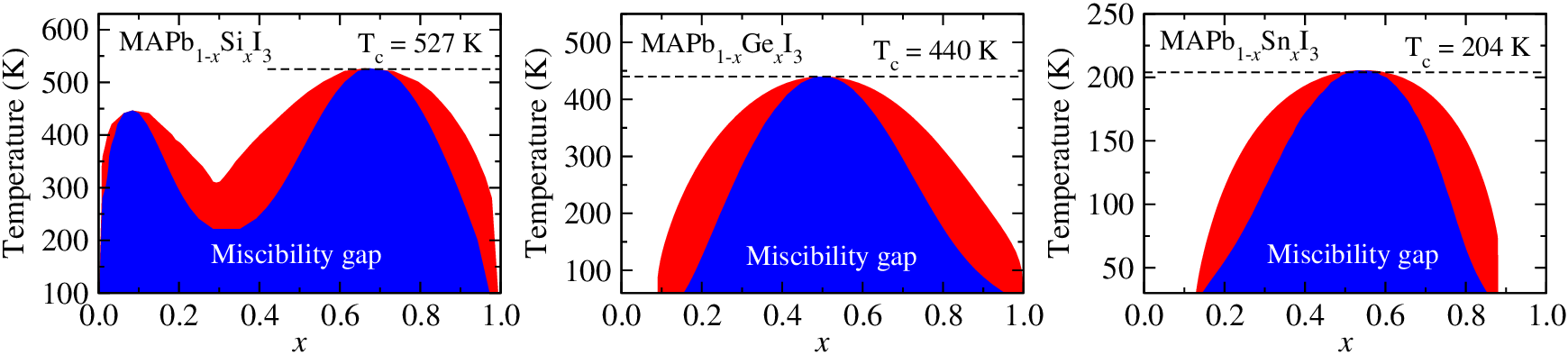}
\caption{Predicted phase diagram of the \ce{MAPb_{1-x}Si_{x}I3}, \ce{MAPb_{1-x}Ge_{x}I3}, 
and \ce{MAPb_{1-x}Sn_{x}I3} alloys at pseudo-cubic structure. The blue and red regions are 
the miscibility gap (spinodal line) and metaestability (defined by the binodal line) regions, 
respectively, while the white region is the stable solid-solution with respect the temperature 
and compositions. 
The dashed line indicates the critical temperature ($T_c$) for each alloy. 
}
\label{fig:phase_diagram}
\end{figure}
To identify regions of stability and metaestability as a function of the temperature
and composition, we built the phase diagram for the alloys at the pseudo-cubic structure, as 
shown in Figure \ref{fig:phase_diagram}. We observe for \ce{MAPb_{1-x}Si_{x}I3} (leftmost), 
\ce{MAPb_{1-x}Ge_{x}I3} (middle), and \ce{MAPb_{1-x}Sn_{x}I3} (rightmost) critical temperatures 
($T_c$) of \num{527}, \num{440}, and \SI{204}{\kelvin}, respectively. Above $T_c$ the solid 
solution are stable for any composition, whereas below them are evidenced the presence of 
miscibility gaps to each alloy defined by spinodal lines, given by $x_1^{\prime}$ and $x_2^{\prime}$ 
(blue regions), as well as binodal lines from the $x_1$ and $x_2$ points defining the $T-x$ 
metaestability regions.

For \ce{MAPb_{1-x}Si_{x}I3}, due to the $\Delta F$ profile observed especially for its formation 
at room temperature (\SI{300}{\kelvin}), as shown in Figure \ref{fig:excess_free}, its phase 
diagram presents unstable regions from \ce{Pb}- to \ce{Si}-rich compositions at low temperatures, 
yielding two miscibility gaps in dependence of the composition region. 
For instance, for \ce{Pb}-rich compositions lie into $x_1^{\prime} = 0.02$ and $x_2^{\prime} = 0.19$, 
whereas the second one, relative to the \ce{Si}-rich compositions, lie to the interval of 
$x_1^{\prime} = 0.45$ up to $x_2^{\prime} = 0.89$. One observes that the first miscibility gap 
reduces as the temperature increases up to \SI{445}{\kelvin}, from which a solid solution is 
formed for \ce{Pb}-rich compositions. However, only from $T_c = \SI{527}{\kelvin}$ at $x = 0.68$ 
the solution solid is stable into all composition interval.
Furthermore, from the end of the first miscibility gap up to the beginning of the second one, i.e., 
between the set of compositions into $x = 0.19-0.45$, the alloy present a metaestable phase resistant 
to the thermal fluctuations due to the valley yielded by the configurations $j$ indicated within 
the dashed retangle in Figure \ref{fig:excess_energy}. 

For the \ce{MAPb_{1-x}Ge_{x}I3} alloy, a stable solid solution is observed in all range of temperatures 
only for \ce{Pb}-rich compositions between $0 < x < x_1$ for $x_1 = 0.20$, while at \SI{300}{\kelvin} 
the miscibility gap appears between $x_1^\prime = 0.31$ and $x_2^\prime = 0.70$. 
At \SI{400}{\kelvin} the intervals for miscibility gap and metaestable phases are shorter than at 
room temperature. 
By comparing the \ce{MAPb_{1-x}Si_{x}I3} and \ce{MAPb_{1-x}Ge_{x}I3} alloys at compositions \ce{Si}-, 
\ce{Ge}-rich, and \ce{Pb}-rich, one observes a behavior very different due to the effect of the 
\ce{Pb}/\ce{Si} and \ce{Pb}/\ce{Ge} metal size differences. Even though there is a stability of 
the \ce{MAPb_{1-x}Ge_{x}I3} alloy for \ce{Pb}-rich into all temperatures, the symmetrical-like 
spinodal line at \SI{300}{\kelvin} yields a stability for a range of \ce{Ge}-rich compositions. 
Additionally, metaestable phases are observed into $x = 0.20-0.31$ and $x = 0.70-0.81$ intervals 
of compositions.

We found that the critical temperatures $T_c$ for \ce{MAPb_{1-x}Si_{x}I3}, \ce{MAPb_{1-x}Ge_{x}I3}, 
and \ce{MAPb_{1-x}Sn_{x}I3}, from which the solid solution at all compositions is stable, correlates 
with the atomic size difference for the metals involved. 
For example, the Figure \ref{fig:phase_diagram} shows also the phase diagram for the \ce{MAPb_{1-x}Sn_{x}I3} 
in which one observes the effects of small difference between the \ce{Pb} and \ce{Sn} atomic size, 
which yields in their high solubility into \ce{MASnI3} and \ce{MAPbI3}, respectively.
Since the critical temperature is $T_c = \SI{204}{\kelvin}$, at \SI{300}{\kelvin} a stable solid solution 
is observed within all range of compositions, which is in agreement with Hao \textit{et al.}\cite{Hao_8094_2014} 
experiment for the synthesis of \ce{MAPb_{1-x}Sn_{x}I3} who observed a high stability of the pseudo-cubic 
structure of the \ce{MAPb_{0.5}Sn_{0.5}I3} alloy, as well as in others compositions.
Furthermore, we found a miscibility gap slightly symmetrical for \ce{MAPb_{1-x}Sn_{x}I3} by appearing 
only at very low temperatures, since the local distortions into the structure are suppressed and the 
entropic effects are restricted to the configurations of the \ce{PbI3^{-}} and \ce{SnI3^{-}} octahedrals. 
Therefore, by taking as reference $T = \SI{443}{\kelvin}$ and $T = \SI{473}{\kelvin}$ as experimental 
temperatures in which the \ce{MAPbI3} and \ce{MASnI3} pure perovskites start to be decomposed\cite{Brunetti_31896_2016,Dimesso_4132_2017} 
our results show that there is a range of temperatures from $T_c = \SI{204}{\kelvin}$ in which the 
\ce{MAPb_{1-x}Sn_{x}I3} is stable as a random alloy before a possible thermal decomposition.
Furthermore, for the others alloys, those results may be as a guide for future synthetic process for 
the \ce{MAPb_{1-x}Si_{x}I3} and \ce{MAPb_{1-x}Ge_{x}I3} alloys, from which it is expected the phase 
segregations for some range of compositions.

\section{Conclusions}

In summary, we have performed first-principles calculations combined with a statistical
approach based on cluster expansion to study the stability, effects of disorder,  
distortions, thermodynamic properties, and phase segregation of the pseudo-cubic phase
of \ce{MAPb_{1-x}B_{x}I3} alloys for \ce{B} = \ce{Si}, \ce{Ge}, and \ce{Sn}.

Our results indicated that the metal atomic size plays an important role on the pseudo-cubic 
properties of the pure perovskites, e.g., as the similar local distortions for the \ce{MAPbI3} 
and \ce{MASnI3} octahedrals since their metals have almost the same atomic size. 
As such, the \ce{MAPb_{1-x}Sn_{x}I3} alloy presents lattice parameters is in good agreement 
with the Vegard’s law for the whole range of compositions, wherein the alloy adopts a random
\ce{PbI3^{-}} and \ce{SnI3^{-}} octahedral configurations.
Conversely, \ce{MASiI3} and \ce{MAGeI3} in pseudo-cubic structure are strongly distorted 
as an effect of their second smaller metal in comparison with \ce{Pb}, by suggesting a higher 
limitation of the organic cation orientation on the lattice directions for the \ce{MAPb_{1-x}Si_{x}I3}
and \ce{MAPb_{1-x}Ge_{x}I3} alloys, since the cavity volume is reduced. 
For those cases, the alloys follow the Vegard’s law for some particular lattice directions, 
whereas the others there is a accentuating bowing throughout the range of compositions.

The thermodynamic results revealed that the \ce{MAPb_{1-x}Ge_{x}I3} alloy is stable 
for \ce{Pb}-rich compositions, i.e., between $0 < x < 0.20$ at \SI{300}{\kelvin}, by 
presenting a preference for an ordered configuration in which one \ce{GeI3^{-}} octahedral 
is surrounded by \ce{PbI3^{-}} octahedrals. Conversely, \ce{MAPb_{1-x}Si_{x}I3} is not 
favored into very large range of compositions, and even though has presented an interval 
of compositions into which the alloy is metaestable (into $x = 0.19-0.45$), it indicated 
a high tendency for segregation phase in \ce{MAPbI3} and \ce{MASiI3} pure perovskites. 
Thus, the addition of small metal atoms yields strong local distortions into the pseudo-cubic 
octahedrals, resulting in very high critical temperatures for these alloys. 
As an exemple of miscibility, the \ce{MAPb_{1-x}Sn_{x}I3} alloy presented a critical 
temperature lower than the room temperature, at \SI{204}{\kelvin}, which is very lower 
than the temperature of decomposition for the \ce{MAPbI3} and \ce{MASnI3} pure perovskites. 
Thus, the alloy is favored as pseudo-cubic as a random alloy in all compositions, revealing 
that there is a safe range of temperatures in which the \ce{MAPb_{1-x}Sn_{x}I3} alloy 
properties can be tuned before the material be thermally decomposed.

Therefore, beyond the temperature as variable, the correlation between composition and 
atomic size, relative to the second metal in \ce{MAPbI3}-based alloys, is a crucial element 
to promotes the phase stability. The thermodynamic characterization of these alloys for 
intermediate \ce{Pb}-based compositions showed the importance of the planning relative to 
the experimental synthesis conditions, such as temperature and composition, aiming the 
structural motifs correlated with their performance into solar cells devices.

\section*{Theoretical Approach and Computational Details}

In this study, to calculate the total energy of the configurations of the alloy
in all the range of compositions, we employed spin-polarized calculations based on
DFT\cite{Hohenberg_B864_1964,Kohn_A1133_1965} within the semilocal Perdew--Burke--Ernzerhof\cite{Perdew_3865_1996}
(PBE) formulation for the exchange-correlation energy functional.
The projected augmented wave\cite{Blochl_17953_1994,Kresse_1758_1999} (PAW) method
as implemented in the Vienna \textit{ab initio} simulation package (VASP), version
$5.4.1$.\cite{Kresse_13115_1993,Kresse_11169_1996} was used to solve the Kohn--Sham (KS) equations,
in which the scalar-relativistic approximation is considered to the core states, as
well the spin-orbit coupling (SOC) interactions.
However, once the SOC is an important relativistic phenomenon in \ce{Pb}-based
perovskites\cite{Even_205301_2012,Even_2999_2013,Even_10161_2015},
especially occurring within non-spherical atomic orbitals and affecting the directionality
of the metal bonds\cite{Manser_12956_2016}, we included SOC interactions also for the
valence states in all our calculations.

For total energy calculations, we employed a plane-waves expansion with kinetic
energy cutoff of \SI{500}{\electronvolt}, by integrating over the Brillouin zone
calculated considering a Monkhorst-Pack \textbf{k}-mesh of $4{\times}4{\times}4$. 
The total energy convergence to \SI{1.0d-5}{\eV} with Gaussian broadening parameter
of \SI{50}{\milli\electronvolt} for all calculations. Finally, the equilibrium of
the Hellmann-Feynman forces on every atom were reached to forces smaller than
\SI{0.010}{\electronvolt/\angstrom}.

\bibliography{jshort_04July2018,boxref_04July2018}

\begin{thebibliography}{79}
\expandafter\ifx\csname natexlab\endcsname\relax\def\natexlab#1{#1}\fi
\expandafter\ifx\csname bibnamefont\endcsname\relax
  \def\bibnamefont#1{#1}\fi
\expandafter\ifx\csname bibfnamefont\endcsname\relax
  \def\bibfnamefont#1{#1}\fi
\expandafter\ifx\csname citenamefont\endcsname\relax
  \def\citenamefont#1{#1}\fi
\expandafter\ifx\csname url\endcsname\relax
  \def\url#1{\texttt{#1}}\fi
\expandafter\ifx\csname urlprefix\endcsname\relax\def\urlprefix{URL }\fi
\providecommand{\bibinfo}[2]{#2}
\providecommand{\eprint}[2][]{\url{#2}}

\bibitem[{\citenamefont{Green et~al.}(2014)\citenamefont{Green, {Ho-Baillie},
  and Snaith}}]{Green_506_2014}
\bibinfo{author}{\bibfnamefont{M.~A.} \bibnamefont{Green}},
  \bibinfo{author}{\bibfnamefont{A.}~\bibnamefont{{Ho-Baillie}}},
  \bibnamefont{and} \bibinfo{author}{\bibfnamefont{H.~J.}
  \bibnamefont{Snaith}}, \bibinfo{journal}{Nat. Photon.}
  \textbf{\bibinfo{volume}{8}}, \bibinfo{pages}{506} (\bibinfo{year}{2014}).

\bibitem[{\citenamefont{Boix et~al.}(2014)\citenamefont{Boix, Nonomura,
  Mathews, and Mhaisalkar}}]{Boix_16_2014}
\bibinfo{author}{\bibfnamefont{P.~P.} \bibnamefont{Boix}},
  \bibinfo{author}{\bibfnamefont{K.}~\bibnamefont{Nonomura}},
  \bibinfo{author}{\bibfnamefont{N.}~\bibnamefont{Mathews}}, \bibnamefont{and}
  \bibinfo{author}{\bibfnamefont{S.~G.} \bibnamefont{Mhaisalkar}},
  \bibinfo{journal}{Materials Today} \textbf{\bibinfo{volume}{17}},
  \bibinfo{pages}{16} (\bibinfo{year}{2014}), ISSN \bibinfo{issn}{1369-7021}.

\bibitem[{\citenamefont{{Gr\"atzel}}(2014)}]{Gratzel_838_2014}
\bibinfo{author}{\bibfnamefont{M.}~\bibnamefont{{Gr\"atzel}}},
  \bibinfo{journal}{Nature} \textbf{\bibinfo{volume}{13}}, \bibinfo{pages}{838}
  (\bibinfo{year}{2014}).

\bibitem[{\citenamefont{Saparov and Mitzi}(2016)}]{Sarapov_4558_2016}
\bibinfo{author}{\bibfnamefont{B.}~\bibnamefont{Saparov}} \bibnamefont{and}
  \bibinfo{author}{\bibfnamefont{D.~B.} \bibnamefont{Mitzi}},
  \bibinfo{journal}{Chem. Rev.} \textbf{\bibinfo{volume}{116}},
  \bibinfo{pages}{4558} (\bibinfo{year}{2016}).

\bibitem[{\citenamefont{Ali et~al.}(2018)\citenamefont{Ali, Hou, Zhu, Yan,
  Zheng, and Su}}]{Ali_718_2018}
\bibinfo{author}{\bibfnamefont{R.}~\bibnamefont{Ali}},
  \bibinfo{author}{\bibfnamefont{G.-J.} \bibnamefont{Hou}},
  \bibinfo{author}{\bibfnamefont{Z.-G.} \bibnamefont{Zhu}},
  \bibinfo{author}{\bibfnamefont{Q.-B.} \bibnamefont{Yan}},
  \bibinfo{author}{\bibfnamefont{Q.-R.} \bibnamefont{Zheng}}, \bibnamefont{and}
  \bibinfo{author}{\bibfnamefont{G.}~\bibnamefont{Su}}, \bibinfo{journal}{Chem.
  Mater.} \textbf{\bibinfo{volume}{30}}, \bibinfo{pages}{718}
  (\bibinfo{year}{2018}).

\bibitem[{\citenamefont{Stranks and Snaith}(2015)}]{Stranks_391_2015}
\bibinfo{author}{\bibfnamefont{S.~D.} \bibnamefont{Stranks}} \bibnamefont{and}
  \bibinfo{author}{\bibfnamefont{H.~J.} \bibnamefont{Snaith}},
  \bibinfo{journal}{Nat. Nanotechnol.} \textbf{\bibinfo{volume}{10}},
  \bibinfo{pages}{391} (\bibinfo{year}{2015}).

\bibitem[{\citenamefont{Manser et~al.}(2016)\citenamefont{Manser, Christians,
  and Kamat}}]{Manser_12956_2016}
\bibinfo{author}{\bibfnamefont{J.~S.} \bibnamefont{Manser}},
  \bibinfo{author}{\bibfnamefont{J.~A.} \bibnamefont{Christians}},
  \bibnamefont{and} \bibinfo{author}{\bibfnamefont{P.~V.} \bibnamefont{Kamat}},
  \bibinfo{journal}{Chem. Rev.} \textbf{\bibinfo{volume}{116}},
  \bibinfo{pages}{12956} (\bibinfo{year}{2016}).

\bibitem[{\citenamefont{Sutherland and Sargent}(2016)}]{Sutherland_295_2016}
\bibinfo{author}{\bibfnamefont{B.~R.} \bibnamefont{Sutherland}}
  \bibnamefont{and} \bibinfo{author}{\bibfnamefont{E.~H.}
  \bibnamefont{Sargent}}, \bibinfo{journal}{Nat. Photon.}
  \textbf{\bibinfo{volume}{10}}, \bibinfo{pages}{295} (\bibinfo{year}{2016}).

\bibitem[{\citenamefont{Cui et~al.}(2015)\citenamefont{Cui, Yuan, Li, Xu, Shen,
  Lin, and Wang}}]{Cui_036004_2015}
\bibinfo{author}{\bibfnamefont{J.}~\bibnamefont{Cui}},
  \bibinfo{author}{\bibfnamefont{H.}~\bibnamefont{Yuan}},
  \bibinfo{author}{\bibfnamefont{J.}~\bibnamefont{Li}},
  \bibinfo{author}{\bibfnamefont{X.}~\bibnamefont{Xu}},
  \bibinfo{author}{\bibfnamefont{Y.}~\bibnamefont{Shen}},
  \bibinfo{author}{\bibfnamefont{H.}~\bibnamefont{Lin}}, \bibnamefont{and}
  \bibinfo{author}{\bibfnamefont{M.}~\bibnamefont{Wang}},
  \bibinfo{journal}{Sci. Technol. Adv. Mater.} \textbf{\bibinfo{volume}{16}},
  \bibinfo{pages}{036004} (\bibinfo{year}{2015}).

\bibitem[{\citenamefont{Gottesman et~al.}(2015)\citenamefont{Gottesman, Gouda,
  Kalanoor, Haltzi, Tirosh, Rosh-Hodesh, Tischler, Zaban, Quarti, Mosconi
  et~al.}}]{Gottesman_2332_2015}
\bibinfo{author}{\bibfnamefont{R.}~\bibnamefont{Gottesman}},
  \bibinfo{author}{\bibfnamefont{L.}~\bibnamefont{Gouda}},
  \bibinfo{author}{\bibfnamefont{B.~S.} \bibnamefont{Kalanoor}},
  \bibinfo{author}{\bibfnamefont{E.}~\bibnamefont{Haltzi}},
  \bibinfo{author}{\bibfnamefont{S.}~\bibnamefont{Tirosh}},
  \bibinfo{author}{\bibfnamefont{E.}~\bibnamefont{Rosh-Hodesh}},
  \bibinfo{author}{\bibfnamefont{Y.}~\bibnamefont{Tischler}},
  \bibinfo{author}{\bibfnamefont{A.}~\bibnamefont{Zaban}},
  \bibinfo{author}{\bibfnamefont{C.}~\bibnamefont{Quarti}},
  \bibinfo{author}{\bibfnamefont{E.}~\bibnamefont{Mosconi}},
  \bibnamefont{et~al.}, \bibinfo{journal}{J. Phys. Chem. Lett.}
  \textbf{\bibinfo{volume}{6}}, \bibinfo{pages}{2332} (\bibinfo{year}{2015}),
  \bibinfo{note}{pMID: 26266613}.

\bibitem[{\citenamefont{Chen et~al.}(2016)\citenamefont{Chen, Zhang, Zhu, Liu,
  Siemens, Gu, Zhu, Shen, Wu, Liao et~al.}}]{Chen_081902_2016}
\bibinfo{author}{\bibfnamefont{Q.}~\bibnamefont{Chen}},
  \bibinfo{author}{\bibfnamefont{C.}~\bibnamefont{Zhang}},
  \bibinfo{author}{\bibfnamefont{M.}~\bibnamefont{Zhu}},
  \bibinfo{author}{\bibfnamefont{S.}~\bibnamefont{Liu}},
  \bibinfo{author}{\bibfnamefont{M.~E.} \bibnamefont{Siemens}},
  \bibinfo{author}{\bibfnamefont{S.}~\bibnamefont{Gu}},
  \bibinfo{author}{\bibfnamefont{J.}~\bibnamefont{Zhu}},
  \bibinfo{author}{\bibfnamefont{J.}~\bibnamefont{Shen}},
  \bibinfo{author}{\bibfnamefont{X.}~\bibnamefont{Wu}},
  \bibinfo{author}{\bibfnamefont{C.}~\bibnamefont{Liao}}, \bibnamefont{et~al.},
  \bibinfo{journal}{Appl. Phys. Lett.} \textbf{\bibinfo{volume}{108}},
  \bibinfo{pages}{081902} (\bibinfo{year}{2016}).

\bibitem[{\citenamefont{Kovalsky et~al.}(2017)\citenamefont{Kovalsky, Wang,
  Marek, Burda, and Dyck}}]{Kovalsky_3228_2017}
\bibinfo{author}{\bibfnamefont{A.}~\bibnamefont{Kovalsky}},
  \bibinfo{author}{\bibfnamefont{L.}~\bibnamefont{Wang}},
  \bibinfo{author}{\bibfnamefont{G.~T.} \bibnamefont{Marek}},
  \bibinfo{author}{\bibfnamefont{C.}~\bibnamefont{Burda}}, \bibnamefont{and}
  \bibinfo{author}{\bibfnamefont{J.~S.} \bibnamefont{Dyck}},
  \bibinfo{journal}{J. Phys. Chem. C} \textbf{\bibinfo{volume}{121}},
  \bibinfo{pages}{3228} (\bibinfo{year}{2017}).

\bibitem[{\citenamefont{Lee et~al.}(2017{\natexlab{a}})\citenamefont{Lee, Kim,
  Bae, Lee, Lin, Yang, and Park}}]{Lee_4270_2017}
\bibinfo{author}{\bibfnamefont{J.-W.} \bibnamefont{Lee}},
  \bibinfo{author}{\bibfnamefont{S.-G.} \bibnamefont{Kim}},
  \bibinfo{author}{\bibfnamefont{S.-H.} \bibnamefont{Bae}},
  \bibinfo{author}{\bibfnamefont{D.-K.} \bibnamefont{Lee}},
  \bibinfo{author}{\bibfnamefont{O.}~\bibnamefont{Lin}},
  \bibinfo{author}{\bibfnamefont{Y.}~\bibnamefont{Yang}}, \bibnamefont{and}
  \bibinfo{author}{\bibfnamefont{N.-G.} \bibnamefont{Park}},
  \bibinfo{journal}{Nano Lett.} \textbf{\bibinfo{volume}{17}},
  \bibinfo{pages}{4270} (\bibinfo{year}{2017}{\natexlab{a}}).

\bibitem[{\citenamefont{Lee et~al.}(2017{\natexlab{b}})\citenamefont{Lee, Li,
  Wong, Zhang, Lai, Yu, Kong, Lin, Urban, Grossman et~al.}}]{Lee_8693_2017}
\bibinfo{author}{\bibfnamefont{W.}~\bibnamefont{Lee}},
  \bibinfo{author}{\bibfnamefont{H.}~\bibnamefont{Li}},
  \bibinfo{author}{\bibfnamefont{A.~B.} \bibnamefont{Wong}},
  \bibinfo{author}{\bibfnamefont{D.}~\bibnamefont{Zhang}},
  \bibinfo{author}{\bibfnamefont{M.}~\bibnamefont{Lai}},
  \bibinfo{author}{\bibfnamefont{Y.}~\bibnamefont{Yu}},
  \bibinfo{author}{\bibfnamefont{Q.}~\bibnamefont{Kong}},
  \bibinfo{author}{\bibfnamefont{E.}~\bibnamefont{Lin}},
  \bibinfo{author}{\bibfnamefont{J.~J.} \bibnamefont{Urban}},
  \bibinfo{author}{\bibfnamefont{J.~C.} \bibnamefont{Grossman}},
  \bibnamefont{et~al.}, \bibinfo{journal}{Proc. Natl. Acad. Sci.}
  \textbf{\bibinfo{volume}{114}}, \bibinfo{pages}{8693}
  (\bibinfo{year}{2017}{\natexlab{b}}), ISSN \bibinfo{issn}{0027-8424}.

\bibitem[{\citenamefont{You et~al.}(2015)\citenamefont{You, Meng, Song, Guo,
  Yang, Chang, Hong, Chen, Zhou, Chen et~al.}}]{You_75_2015}
\bibinfo{author}{\bibfnamefont{J.}~\bibnamefont{You}},
  \bibinfo{author}{\bibfnamefont{L.}~\bibnamefont{Meng}},
  \bibinfo{author}{\bibfnamefont{T.}~\bibnamefont{Song}},
  \bibinfo{author}{\bibfnamefont{T.}~\bibnamefont{Guo}},
  \bibinfo{author}{\bibfnamefont{Y.~M.} \bibnamefont{Yang}},
  \bibinfo{author}{\bibfnamefont{W.}~\bibnamefont{Chang}},
  \bibinfo{author}{\bibfnamefont{Z.}~\bibnamefont{Hong}},
  \bibinfo{author}{\bibfnamefont{H.}~\bibnamefont{Chen}},
  \bibinfo{author}{\bibfnamefont{H.}~\bibnamefont{Zhou}},
  \bibinfo{author}{\bibfnamefont{Q.}~\bibnamefont{Chen}}, \bibnamefont{et~al.},
  \bibinfo{journal}{Nat. Nanotechnol.} \textbf{\bibinfo{volume}{11}},
  \bibinfo{pages}{75} (\bibinfo{year}{2015}).

\bibitem[{\citenamefont{Raza et~al.}(2018)\citenamefont{Raza, Aziz, and
  Ahmad}}]{Raza_20952_2018}
\bibinfo{author}{\bibfnamefont{E.}~\bibnamefont{Raza}},
  \bibinfo{author}{\bibfnamefont{F.}~\bibnamefont{Aziz}}, \bibnamefont{and}
  \bibinfo{author}{\bibfnamefont{Z.}~\bibnamefont{Ahmad}},
  \bibinfo{journal}{RSC Adv.} \textbf{\bibinfo{volume}{8}},
  \bibinfo{pages}{20952} (\bibinfo{year}{2018}).

\bibitem[{\citenamefont{Abdelmageed et~al.}(2018)\citenamefont{Abdelmageed,
  Sully, Bonabi~Naghadeh, {El-Hag}~Ali, Carter, and
  Zhang}}]{Abdelmageed_387_2018}
\bibinfo{author}{\bibfnamefont{G.}~\bibnamefont{Abdelmageed}},
  \bibinfo{author}{\bibfnamefont{H.~R.} \bibnamefont{Sully}},
  \bibinfo{author}{\bibfnamefont{S.}~\bibnamefont{Bonabi~Naghadeh}},
  \bibinfo{author}{\bibfnamefont{A.}~\bibnamefont{{El-Hag}~Ali}},
  \bibinfo{author}{\bibfnamefont{S.~A.} \bibnamefont{Carter}},
  \bibnamefont{and} \bibinfo{author}{\bibfnamefont{J.~Z.} \bibnamefont{Zhang}},
  \bibinfo{journal}{ACS Appl. Energy Mater.} \textbf{\bibinfo{volume}{1}},
  \bibinfo{pages}{387} (\bibinfo{year}{2018}).

\bibitem[{\citenamefont{Baikie et~al.}(2013)\citenamefont{Baikie, Fang, Kadro,
  Schreyer, Wei, Mhaisalkar, Graetzel, and White}}]{Baikie_5628_2013}
\bibinfo{author}{\bibfnamefont{T.}~\bibnamefont{Baikie}},
  \bibinfo{author}{\bibfnamefont{Y.}~\bibnamefont{Fang}},
  \bibinfo{author}{\bibfnamefont{J.~M.} \bibnamefont{Kadro}},
  \bibinfo{author}{\bibfnamefont{M.}~\bibnamefont{Schreyer}},
  \bibinfo{author}{\bibfnamefont{F.}~\bibnamefont{Wei}},
  \bibinfo{author}{\bibfnamefont{S.~G.} \bibnamefont{Mhaisalkar}},
  \bibinfo{author}{\bibfnamefont{M.}~\bibnamefont{Graetzel}}, \bibnamefont{and}
  \bibinfo{author}{\bibfnamefont{T.~J.} \bibnamefont{White}},
  \bibinfo{journal}{J. Mater. Chem. A} \textbf{\bibinfo{volume}{1}},
  \bibinfo{pages}{5628} (\bibinfo{year}{2013}).

\bibitem[{\citenamefont{G et~al.}(2016)\citenamefont{G, Mahale, Kore,
  Mukherjee, Pavan, De, Ghara, Sundaresan, Pandey, Guru~Row
  et~al.}}]{Sharada_2412_2016}
\bibinfo{author}{\bibfnamefont{S.}~\bibnamefont{G}},
  \bibinfo{author}{\bibfnamefont{P.}~\bibnamefont{Mahale}},
  \bibinfo{author}{\bibfnamefont{B.~P.} \bibnamefont{Kore}},
  \bibinfo{author}{\bibfnamefont{S.}~\bibnamefont{Mukherjee}},
  \bibinfo{author}{\bibfnamefont{M.~S.} \bibnamefont{Pavan}},
  \bibinfo{author}{\bibfnamefont{C.}~\bibnamefont{De}},
  \bibinfo{author}{\bibfnamefont{S.}~\bibnamefont{Ghara}},
  \bibinfo{author}{\bibfnamefont{A.}~\bibnamefont{Sundaresan}},
  \bibinfo{author}{\bibfnamefont{A.}~\bibnamefont{Pandey}},
  \bibinfo{author}{\bibfnamefont{T.~N.} \bibnamefont{Guru~Row}},
  \bibnamefont{et~al.}, \bibinfo{journal}{J. Phys. Chem. Lett.}
  \textbf{\bibinfo{volume}{7}}, \bibinfo{pages}{2412} (\bibinfo{year}{2016}).

\bibitem[{\citenamefont{Stoumpos et~al.}(2013)\citenamefont{Stoumpos,
  Malliakas, and Kanatzidis}}]{Stoumpos_9019_2013}
\bibinfo{author}{\bibfnamefont{C.~C.} \bibnamefont{Stoumpos}},
  \bibinfo{author}{\bibfnamefont{C.~D.} \bibnamefont{Malliakas}},
  \bibnamefont{and} \bibinfo{author}{\bibfnamefont{M.~G.}
  \bibnamefont{Kanatzidis}}, \bibinfo{journal}{Inorg. Chem.}
  \textbf{\bibinfo{volume}{52}}, \bibinfo{pages}{9019} (\bibinfo{year}{2013}).

\bibitem[{\citenamefont{Qinglong et~al.}(2015)\citenamefont{Qinglong, Dominic,
  Jue, L., Chong, and Tao}}]{Qinglong_7617_2015}
\bibinfo{author}{\bibfnamefont{J.}~\bibnamefont{Qinglong}},
  \bibinfo{author}{\bibfnamefont{R.}~\bibnamefont{Dominic}},
  \bibinfo{author}{\bibfnamefont{G.}~\bibnamefont{Jue}},
  \bibinfo{author}{\bibfnamefont{P.~E.} \bibnamefont{L.}},
  \bibinfo{author}{\bibfnamefont{Z.}~\bibnamefont{Chong}}, \bibnamefont{and}
  \bibinfo{author}{\bibfnamefont{X.}~\bibnamefont{Tao}},
  \bibinfo{journal}{Angew. Chem. Int. Ed.} \textbf{\bibinfo{volume}{54}},
  \bibinfo{pages}{7617} (\bibinfo{year}{2015}).

\bibitem[{\citenamefont{Joseph et~al.}(2015)\citenamefont{Joseph, Tonio, A.,
  Gary, Leeor, {Yueh-Lin}, Igor, R., Yitzhak, S. et~al.}}]{Joseph_5102_2015}
\bibinfo{author}{\bibfnamefont{B.}~\bibnamefont{Joseph}},
  \bibinfo{author}{\bibfnamefont{B.}~\bibnamefont{Tonio}},
  \bibinfo{author}{\bibfnamefont{E.~D.} \bibnamefont{A.}},
  \bibinfo{author}{\bibfnamefont{H.}~\bibnamefont{Gary}},
  \bibinfo{author}{\bibfnamefont{K.}~\bibnamefont{Leeor}},
  \bibinfo{author}{\bibfnamefont{L.}~\bibnamefont{{Yueh-Lin}}},
  \bibinfo{author}{\bibfnamefont{L.}~\bibnamefont{Igor}},
  \bibinfo{author}{\bibfnamefont{M.~S.} \bibnamefont{R.}},
  \bibinfo{author}{\bibfnamefont{M.}~\bibnamefont{Yitzhak}},
  \bibinfo{author}{\bibfnamefont{M.~J.} \bibnamefont{S.}},
  \bibnamefont{et~al.}, \bibinfo{journal}{Adv. Mater.}
  \textbf{\bibinfo{volume}{27}}, \bibinfo{pages}{5102} (\bibinfo{year}{2015}).

\bibitem[{\citenamefont{Lee et~al.}(2016)\citenamefont{Lee, Kim, Bae, Cho,
  Chung, Mundt, Lee, Park, Park, Schubert et~al.}}]{Lee_1_2016}
\bibinfo{author}{\bibfnamefont{S.-W.} \bibnamefont{Lee}},
  \bibinfo{author}{\bibfnamefont{S.}~\bibnamefont{Kim}},
  \bibinfo{author}{\bibfnamefont{S.}~\bibnamefont{Bae}},
  \bibinfo{author}{\bibfnamefont{K.}~\bibnamefont{Cho}},
  \bibinfo{author}{\bibfnamefont{T.}~\bibnamefont{Chung}},
  \bibinfo{author}{\bibfnamefont{L.~E.} \bibnamefont{Mundt}},
  \bibinfo{author}{\bibfnamefont{S.}~\bibnamefont{Lee}},
  \bibinfo{author}{\bibfnamefont{S.}~\bibnamefont{Park}},
  \bibinfo{author}{\bibfnamefont{H.}~\bibnamefont{Park}},
  \bibinfo{author}{\bibfnamefont{M.~C.} \bibnamefont{Schubert}},
  \bibnamefont{et~al.}, \bibinfo{journal}{Sci. Rep.}
  \textbf{\bibinfo{volume}{6}}, \bibinfo{pages}{1} (\bibinfo{year}{2016}).

\bibitem[{\citenamefont{Zhang et~al.}(2018)\citenamefont{Zhang, Chen, Xu,
  Xiang, Gong, Walsh, and Wei}}]{Zhang_036104_2018}
\bibinfo{author}{\bibfnamefont{Y.-Y.} \bibnamefont{Zhang}},
  \bibinfo{author}{\bibfnamefont{S.}~\bibnamefont{Chen}},
  \bibinfo{author}{\bibfnamefont{P.}~\bibnamefont{Xu}},
  \bibinfo{author}{\bibfnamefont{H.}~\bibnamefont{Xiang}},
  \bibinfo{author}{\bibfnamefont{X.-G.} \bibnamefont{Gong}},
  \bibinfo{author}{\bibfnamefont{A.}~\bibnamefont{Walsh}}, \bibnamefont{and}
  \bibinfo{author}{\bibfnamefont{S.-H.} \bibnamefont{Wei}},
  \bibinfo{journal}{Chinese Physics Letters} \textbf{\bibinfo{volume}{35}},
  \bibinfo{pages}{036104} (\bibinfo{year}{2018}).

\bibitem[{\citenamefont{Brunetti et~al.}(2016)\citenamefont{Brunetti, Cavallo,
  Ciccioli, Gigli, and Latini}}]{Brunetti_31896_2016}
\bibinfo{author}{\bibfnamefont{B.}~\bibnamefont{Brunetti}},
  \bibinfo{author}{\bibfnamefont{C.}~\bibnamefont{Cavallo}},
  \bibinfo{author}{\bibfnamefont{A.}~\bibnamefont{Ciccioli}},
  \bibinfo{author}{\bibfnamefont{G.}~\bibnamefont{Gigli}}, \bibnamefont{and}
  \bibinfo{author}{\bibfnamefont{A.}~\bibnamefont{Latini}},
  \bibinfo{journal}{Sci. Rep.} \textbf{\bibinfo{volume}{6}},
  \bibinfo{pages}{31896} (\bibinfo{year}{2016}).

\bibitem[{\citenamefont{Dimesso et~al.}(2017)\citenamefont{Dimesso, St{\"o}hr,
  Das, Mayer, and Jaegermann}}]{Dimesso_4132_2017}
\bibinfo{author}{\bibfnamefont{L.}~\bibnamefont{Dimesso}},
  \bibinfo{author}{\bibfnamefont{M.}~\bibnamefont{St{\"o}hr}},
  \bibinfo{author}{\bibfnamefont{C.}~\bibnamefont{Das}},
  \bibinfo{author}{\bibfnamefont{T.}~\bibnamefont{Mayer}}, \bibnamefont{and}
  \bibinfo{author}{\bibfnamefont{W.}~\bibnamefont{Jaegermann}},
  \bibinfo{journal}{J. Mater. Res.} \textbf{\bibinfo{volume}{32}},
  \bibinfo{pages}{4132} (\bibinfo{year}{2017}).

\bibitem[{\citenamefont{Norman et~al.}(2014)\citenamefont{Norman, Peng,
  Giuliano, Tae-Youl, K., Joachim, and Michael}}]{Norman_3151_2015}
\bibinfo{author}{\bibfnamefont{P.}~\bibnamefont{Norman}},
  \bibinfo{author}{\bibfnamefont{G.}~\bibnamefont{Peng}},
  \bibinfo{author}{\bibfnamefont{G.}~\bibnamefont{Giuliano}},
  \bibinfo{author}{\bibfnamefont{Y.}~\bibnamefont{Tae-Youl}},
  \bibinfo{author}{\bibfnamefont{N.~M.} \bibnamefont{K.}},
  \bibinfo{author}{\bibfnamefont{M.}~\bibnamefont{Joachim}}, \bibnamefont{and}
  \bibinfo{author}{\bibfnamefont{G.}~\bibnamefont{Michael}},
  \bibinfo{journal}{Angew. Chem. Int. Ed.} \textbf{\bibinfo{volume}{53}},
  \bibinfo{pages}{3151} (\bibinfo{year}{2014}).

\bibitem[{\citenamefont{Weber et~al.}(2016)\citenamefont{Weber, Charles, and
  Weller}}]{Weber_15375_2016}
\bibinfo{author}{\bibfnamefont{O.~J.} \bibnamefont{Weber}},
  \bibinfo{author}{\bibfnamefont{B.}~\bibnamefont{Charles}}, \bibnamefont{and}
  \bibinfo{author}{\bibfnamefont{M.~T.} \bibnamefont{Weller}},
  \bibinfo{journal}{J. Mater. Chem. A} \textbf{\bibinfo{volume}{4}},
  \bibinfo{pages}{15375} (\bibinfo{year}{2016}).

\bibitem[{\citenamefont{Kubicki et~al.}(2017)\citenamefont{Kubicki, Prochowicz,
  Hofstetter, Zakeeruddin, Grätzel, and Emsley}}]{Kubicki_14173_2017}
\bibinfo{author}{\bibfnamefont{D.~J.} \bibnamefont{Kubicki}},
  \bibinfo{author}{\bibfnamefont{D.}~\bibnamefont{Prochowicz}},
  \bibinfo{author}{\bibfnamefont{A.}~\bibnamefont{Hofstetter}},
  \bibinfo{author}{\bibfnamefont{S.~M.} \bibnamefont{Zakeeruddin}},
  \bibinfo{author}{\bibfnamefont{M.}~\bibnamefont{Grätzel}}, \bibnamefont{and}
  \bibinfo{author}{\bibfnamefont{L.}~\bibnamefont{Emsley}},
  \bibinfo{journal}{J. Am. Chem. Soc.} \textbf{\bibinfo{volume}{139}},
  \bibinfo{pages}{14173} (\bibinfo{year}{2017}).

\bibitem[{\citenamefont{Ogomi et~al.}(2014)\citenamefont{Ogomi, Morita,
  Tsukamoto, Saitho, Fujikawa, Shen, Toyoda, Yoshino, Pandey, Ma
  et~al.}}]{Ogomi_1004_2014}
\bibinfo{author}{\bibfnamefont{Y.}~\bibnamefont{Ogomi}},
  \bibinfo{author}{\bibfnamefont{A.}~\bibnamefont{Morita}},
  \bibinfo{author}{\bibfnamefont{S.}~\bibnamefont{Tsukamoto}},
  \bibinfo{author}{\bibfnamefont{T.}~\bibnamefont{Saitho}},
  \bibinfo{author}{\bibfnamefont{N.}~\bibnamefont{Fujikawa}},
  \bibinfo{author}{\bibfnamefont{Q.}~\bibnamefont{Shen}},
  \bibinfo{author}{\bibfnamefont{T.}~\bibnamefont{Toyoda}},
  \bibinfo{author}{\bibfnamefont{K.}~\bibnamefont{Yoshino}},
  \bibinfo{author}{\bibfnamefont{S.~S.} \bibnamefont{Pandey}},
  \bibinfo{author}{\bibfnamefont{T.}~\bibnamefont{Ma}}, \bibnamefont{et~al.},
  \bibinfo{journal}{J. Phys. Chem. Lett.} \textbf{\bibinfo{volume}{5}},
  \bibinfo{pages}{1004} (\bibinfo{year}{2014}), \bibinfo{note}{pMID: 26270980}.

\bibitem[{\citenamefont{Hao et~al.}(2014)\citenamefont{Hao, Stoumpos, Chang,
  and Kanatzidis}}]{Hao_8094_2014}
\bibinfo{author}{\bibfnamefont{F.}~\bibnamefont{Hao}},
  \bibinfo{author}{\bibfnamefont{C.~C.} \bibnamefont{Stoumpos}},
  \bibinfo{author}{\bibfnamefont{R.~P.~H.} \bibnamefont{Chang}},
  \bibnamefont{and} \bibinfo{author}{\bibfnamefont{M.~G.}
  \bibnamefont{Kanatzidis}}, \bibinfo{journal}{J. Am. Chem. Soc.}
  \textbf{\bibinfo{volume}{136}}, \bibinfo{pages}{8094} (\bibinfo{year}{2014}).

\bibitem[{\citenamefont{Feng et~al.}(2015)\citenamefont{Feng, Paudel, Tsymbal,
  and Zeng}}]{Feng_8227_2015}
\bibinfo{author}{\bibfnamefont{H.-J.} \bibnamefont{Feng}},
  \bibinfo{author}{\bibfnamefont{T.~R.} \bibnamefont{Paudel}},
  \bibinfo{author}{\bibfnamefont{E.~Y.} \bibnamefont{Tsymbal}},
  \bibnamefont{and} \bibinfo{author}{\bibfnamefont{X.~C.} \bibnamefont{Zeng}},
  \bibinfo{journal}{J. Am. Chem. Soc.} \textbf{\bibinfo{volume}{137}},
  \bibinfo{pages}{8227} (\bibinfo{year}{2015}).

\bibitem[{\citenamefont{Sampson et~al.}(2017)\citenamefont{Sampson, Park,
  Schaller, Chan, and Martinson}}]{Sampson_3578_2017}
\bibinfo{author}{\bibfnamefont{M.~D.} \bibnamefont{Sampson}},
  \bibinfo{author}{\bibfnamefont{J.~S.} \bibnamefont{Park}},
  \bibinfo{author}{\bibfnamefont{R.~D.} \bibnamefont{Schaller}},
  \bibinfo{author}{\bibfnamefont{M.~K.~Y.} \bibnamefont{Chan}},
  \bibnamefont{and} \bibinfo{author}{\bibfnamefont{A.~B.~F.}
  \bibnamefont{Martinson}}, \bibinfo{journal}{J. Mater. Chem. A}
  \textbf{\bibinfo{volume}{5}}, \bibinfo{pages}{3578} (\bibinfo{year}{2017}).

\bibitem[{\citenamefont{Colella et~al.}(2013)\citenamefont{Colella, Mosconi,
  Fedeli, Listorti, Gazza, Orlandi, Ferro, Besagni, Rizzo, Calestani
  et~al.}}]{Colella_4613_2013}
\bibinfo{author}{\bibfnamefont{S.}~\bibnamefont{Colella}},
  \bibinfo{author}{\bibfnamefont{E.}~\bibnamefont{Mosconi}},
  \bibinfo{author}{\bibfnamefont{P.}~\bibnamefont{Fedeli}},
  \bibinfo{author}{\bibfnamefont{A.}~\bibnamefont{Listorti}},
  \bibinfo{author}{\bibfnamefont{F.}~\bibnamefont{Gazza}},
  \bibinfo{author}{\bibfnamefont{F.}~\bibnamefont{Orlandi}},
  \bibinfo{author}{\bibfnamefont{P.}~\bibnamefont{Ferro}},
  \bibinfo{author}{\bibfnamefont{T.}~\bibnamefont{Besagni}},
  \bibinfo{author}{\bibfnamefont{A.}~\bibnamefont{Rizzo}},
  \bibinfo{author}{\bibfnamefont{G.}~\bibnamefont{Calestani}},
  \bibnamefont{et~al.}, \bibinfo{journal}{Chem. Mater.}
  \textbf{\bibinfo{volume}{25}}, \bibinfo{pages}{4613} (\bibinfo{year}{2013}).

\bibitem[{\citenamefont{Pathak et~al.}(2015)\citenamefont{Pathak, Sakai,
  Wisnivesky Rocca~Rivarola, Stranks, Liu, Eperon, Ducati, Wojciechowski,
  Griffiths, Haghighirad et~al.}}]{Pathak_8066_2015}
\bibinfo{author}{\bibfnamefont{S.}~\bibnamefont{Pathak}},
  \bibinfo{author}{\bibfnamefont{N.}~\bibnamefont{Sakai}},
  \bibinfo{author}{\bibfnamefont{F.}~\bibnamefont{Wisnivesky Rocca~Rivarola}},
  \bibinfo{author}{\bibfnamefont{S.~D.} \bibnamefont{Stranks}},
  \bibinfo{author}{\bibfnamefont{J.}~\bibnamefont{Liu}},
  \bibinfo{author}{\bibfnamefont{G.~E.} \bibnamefont{Eperon}},
  \bibinfo{author}{\bibfnamefont{C.}~\bibnamefont{Ducati}},
  \bibinfo{author}{\bibfnamefont{K.}~\bibnamefont{Wojciechowski}},
  \bibinfo{author}{\bibfnamefont{J.~T.} \bibnamefont{Griffiths}},
  \bibinfo{author}{\bibfnamefont{A.~A.} \bibnamefont{Haghighirad}},
  \bibnamefont{et~al.}, \bibinfo{journal}{Chem. Mater.}
  \textbf{\bibinfo{volume}{27}}, \bibinfo{pages}{8066} (\bibinfo{year}{2015}).

\bibitem[{\citenamefont{Zejiao et~al.}(2017)\citenamefont{Zejiao, Jia, Yonghua,
  Qi, Yufeng, Haijuan, Yingdong, and Wei}}]{Shi_1605005_2017}
\bibinfo{author}{\bibfnamefont{S.}~\bibnamefont{Zejiao}},
  \bibinfo{author}{\bibfnamefont{G.}~\bibnamefont{Jia}},
  \bibinfo{author}{\bibfnamefont{C.}~\bibnamefont{Yonghua}},
  \bibinfo{author}{\bibfnamefont{L.}~\bibnamefont{Qi}},
  \bibinfo{author}{\bibfnamefont{P.}~\bibnamefont{Yufeng}},
  \bibinfo{author}{\bibfnamefont{Z.}~\bibnamefont{Haijuan}},
  \bibinfo{author}{\bibfnamefont{X.}~\bibnamefont{Yingdong}}, \bibnamefont{and}
  \bibinfo{author}{\bibfnamefont{H.}~\bibnamefont{Wei}},
  \bibinfo{journal}{Advanced Materials} \textbf{\bibinfo{volume}{29}},
  \bibinfo{pages}{1605005} (\bibinfo{year}{2017}).

\bibitem[{\citenamefont{Hemant et~al.}(2014)\citenamefont{Hemant, Sabba, Lin,
  P., Ramanujam, Tom, Chen, Hong, Ramamoorthy, Mark et~al.}}]{Memant_7122_2014}
\bibinfo{author}{\bibfnamefont{K.~M.} \bibnamefont{Hemant}},
  \bibinfo{author}{\bibfnamefont{D.}~\bibnamefont{Sabba}},
  \bibinfo{author}{\bibfnamefont{L.~W.} \bibnamefont{Lin}},
  \bibinfo{author}{\bibfnamefont{B.~P.} \bibnamefont{P.}},
  \bibinfo{author}{\bibfnamefont{P.~R.} \bibnamefont{Ramanujam}},
  \bibinfo{author}{\bibfnamefont{B.}~\bibnamefont{Tom}},
  \bibinfo{author}{\bibfnamefont{S.}~\bibnamefont{Chen}},
  \bibinfo{author}{\bibfnamefont{D.}~\bibnamefont{Hong}},
  \bibinfo{author}{\bibfnamefont{R.}~\bibnamefont{Ramamoorthy}},
  \bibinfo{author}{\bibfnamefont{A.}~\bibnamefont{Mark}}, \bibnamefont{et~al.},
  \bibinfo{journal}{Adv. Mater.} \textbf{\bibinfo{volume}{26}},
  \bibinfo{pages}{7122} (\bibinfo{year}{2014}).

\bibitem[{\citenamefont{Cortecchia et~al.}(2016)\citenamefont{Cortecchia, Dewi,
  Yin, Bruno, Chen, Baikie, Boix, {Gr\"atzel}, Mhaisalkar, Soci
  et~al.}}]{Cortecchia_1044_2016}
\bibinfo{author}{\bibfnamefont{D.}~\bibnamefont{Cortecchia}},
  \bibinfo{author}{\bibfnamefont{H.~A.} \bibnamefont{Dewi}},
  \bibinfo{author}{\bibfnamefont{J.}~\bibnamefont{Yin}},
  \bibinfo{author}{\bibfnamefont{A.}~\bibnamefont{Bruno}},
  \bibinfo{author}{\bibfnamefont{S.}~\bibnamefont{Chen}},
  \bibinfo{author}{\bibfnamefont{T.}~\bibnamefont{Baikie}},
  \bibinfo{author}{\bibfnamefont{P.~P.} \bibnamefont{Boix}},
  \bibinfo{author}{\bibfnamefont{M.}~\bibnamefont{{Gr\"atzel}}},
  \bibinfo{author}{\bibfnamefont{S.}~\bibnamefont{Mhaisalkar}},
  \bibinfo{author}{\bibfnamefont{C.}~\bibnamefont{Soci}}, \bibnamefont{et~al.},
  \bibinfo{journal}{Inorg. Chem.} \textbf{\bibinfo{volume}{55}},
  \bibinfo{pages}{1044} (\bibinfo{year}{2016}).

\bibitem[{\citenamefont{Zhang et~al.}(2016)\citenamefont{Zhang, Shang, Wang,
  Huang, Xu, Hu, Zhu, and Han}}]{Zhang_535_2016}
\bibinfo{author}{\bibfnamefont{J.}~\bibnamefont{Zhang}},
  \bibinfo{author}{\bibfnamefont{M.-h.} \bibnamefont{Shang}},
  \bibinfo{author}{\bibfnamefont{P.}~\bibnamefont{Wang}},
  \bibinfo{author}{\bibfnamefont{X.}~\bibnamefont{Huang}},
  \bibinfo{author}{\bibfnamefont{J.}~\bibnamefont{Xu}},
  \bibinfo{author}{\bibfnamefont{Z.}~\bibnamefont{Hu}},
  \bibinfo{author}{\bibfnamefont{Y.}~\bibnamefont{Zhu}}, \bibnamefont{and}
  \bibinfo{author}{\bibfnamefont{L.}~\bibnamefont{Han}}, \bibinfo{journal}{ACS
  Energy Lett.} \textbf{\bibinfo{volume}{1}}, \bibinfo{pages}{535}
  (\bibinfo{year}{2016}).

\bibitem[{\citenamefont{Yang et~al.}(2017)\citenamefont{Yang, Lv, Zhao, Xu, Fu,
  Zhan, Zunger, and Zhang}}]{Yang_524_2017}
\bibinfo{author}{\bibfnamefont{D.}~\bibnamefont{Yang}},
  \bibinfo{author}{\bibfnamefont{J.}~\bibnamefont{Lv}},
  \bibinfo{author}{\bibfnamefont{X.}~\bibnamefont{Zhao}},
  \bibinfo{author}{\bibfnamefont{Q.}~\bibnamefont{Xu}},
  \bibinfo{author}{\bibfnamefont{Y.}~\bibnamefont{Fu}},
  \bibinfo{author}{\bibfnamefont{Y.}~\bibnamefont{Zhan}},
  \bibinfo{author}{\bibfnamefont{A.}~\bibnamefont{Zunger}}, \bibnamefont{and}
  \bibinfo{author}{\bibfnamefont{L.}~\bibnamefont{Zhang}},
  \bibinfo{journal}{Chem. Mater.} \textbf{\bibinfo{volume}{29}},
  \bibinfo{pages}{524} (\bibinfo{year}{2017}).

\bibitem[{\citenamefont{Kopacic et~al.}(2018)\citenamefont{Kopacic,
  Friesenbichler, Hoefler, Kunert, Plank, Rath, and
  Trimmel}}]{Kopacic_343_2018}
\bibinfo{author}{\bibfnamefont{I.}~\bibnamefont{Kopacic}},
  \bibinfo{author}{\bibfnamefont{B.}~\bibnamefont{Friesenbichler}},
  \bibinfo{author}{\bibfnamefont{S.~F.} \bibnamefont{Hoefler}},
  \bibinfo{author}{\bibfnamefont{B.}~\bibnamefont{Kunert}},
  \bibinfo{author}{\bibfnamefont{H.}~\bibnamefont{Plank}},
  \bibinfo{author}{\bibfnamefont{T.}~\bibnamefont{Rath}}, \bibnamefont{and}
  \bibinfo{author}{\bibfnamefont{G.}~\bibnamefont{Trimmel}},
  \bibinfo{journal}{ACS Appl. Energy Mater.} \textbf{\bibinfo{volume}{1}},
  \bibinfo{pages}{343} (\bibinfo{year}{2018}),
  \eprint{https://doi.org/10.1021/acsaem.8b00007},
  \urlprefix\url{https://doi.org/10.1021/acsaem.8b00007}.

\bibitem[{\citenamefont{Hu et~al.}(2017)\citenamefont{Hu, Dong, and
  Zhang}}]{Hu_11436_2017}
\bibinfo{author}{\bibfnamefont{H.}~\bibnamefont{Hu}},
  \bibinfo{author}{\bibfnamefont{B.}~\bibnamefont{Dong}}, \bibnamefont{and}
  \bibinfo{author}{\bibfnamefont{W.}~\bibnamefont{Zhang}}, \bibinfo{journal}{J.
  Mater. Chem. A} \textbf{\bibinfo{volume}{5}}, \bibinfo{pages}{11436}
  (\bibinfo{year}{2017}).

\bibitem[{\citenamefont{Stoumpos et~al.}(2015)\citenamefont{Stoumpos, Frazer,
  Clark, Kim, Rhim, Freeman, Ketterson, Jang, and
  Kanatzidis}}]{Stoumpos_6804_2015}
\bibinfo{author}{\bibfnamefont{C.~C.} \bibnamefont{Stoumpos}},
  \bibinfo{author}{\bibfnamefont{L.}~\bibnamefont{Frazer}},
  \bibinfo{author}{\bibfnamefont{D.~J.} \bibnamefont{Clark}},
  \bibinfo{author}{\bibfnamefont{Y.~S.} \bibnamefont{Kim}},
  \bibinfo{author}{\bibfnamefont{S.~H.} \bibnamefont{Rhim}},
  \bibinfo{author}{\bibfnamefont{A.~J.} \bibnamefont{Freeman}},
  \bibinfo{author}{\bibfnamefont{J.~B.} \bibnamefont{Ketterson}},
  \bibinfo{author}{\bibfnamefont{J.~I.} \bibnamefont{Jang}}, \bibnamefont{and}
  \bibinfo{author}{\bibfnamefont{M.~G.} \bibnamefont{Kanatzidis}},
  \bibinfo{journal}{J. Am. Chem. Soc.} \textbf{\bibinfo{volume}{137}},
  \bibinfo{pages}{6804} (\bibinfo{year}{2015}).

\bibitem[{\citenamefont{Sun et~al.}(2016)\citenamefont{Sun, Li, Feng, and
  Li}}]{Sun_14408_2016}
\bibinfo{author}{\bibfnamefont{P.-P.} \bibnamefont{Sun}},
  \bibinfo{author}{\bibfnamefont{Q.-S.} \bibnamefont{Li}},
  \bibinfo{author}{\bibfnamefont{S.}~\bibnamefont{Feng}}, \bibnamefont{and}
  \bibinfo{author}{\bibfnamefont{Z.-S.} \bibnamefont{Li}},
  \bibinfo{journal}{Phys. Chem. Chem. Phys.} \textbf{\bibinfo{volume}{18}},
  \bibinfo{pages}{14408} (\bibinfo{year}{2016}).

\bibitem[{\citenamefont{Goldschmidt}(1926)}]{Goldschmidt_477_1926}
\bibinfo{author}{\bibfnamefont{V.~M.} \bibnamefont{Goldschmidt}},
  \bibinfo{journal}{Naturwissenschaften} \textbf{\bibinfo{volume}{14}},
  \bibinfo{pages}{477} (\bibinfo{year}{1926}), ISSN \bibinfo{issn}{1432-1904}.

\bibitem[{\citenamefont{Kieslich et~al.}(2015)\citenamefont{Kieslich, Sun, and
  Cheetham}}]{Kieslich_3430_2015}
\bibinfo{author}{\bibfnamefont{G.}~\bibnamefont{Kieslich}},
  \bibinfo{author}{\bibfnamefont{S.}~\bibnamefont{Sun}}, \bibnamefont{and}
  \bibinfo{author}{\bibfnamefont{A.~K.} \bibnamefont{Cheetham}},
  \bibinfo{journal}{Chem. Sci.} \textbf{\bibinfo{volume}{6}},
  \bibinfo{pages}{3430} (\bibinfo{year}{2015}).

\bibitem[{\citenamefont{Stoumpos and Kanatzidis}(2015)}]{Stoumpos_2791_2015}
\bibinfo{author}{\bibfnamefont{C.~C.} \bibnamefont{Stoumpos}} \bibnamefont{and}
  \bibinfo{author}{\bibfnamefont{M.~G.} \bibnamefont{Kanatzidis}},
  \bibinfo{journal}{Acc. Chem. Res.} \textbf{\bibinfo{volume}{48}},
  \bibinfo{pages}{2791} (\bibinfo{year}{2015}), \bibinfo{note}{pMID: 26350149}.

\bibitem[{\citenamefont{Isarov et~al.}(2017)\citenamefont{Isarov, Tan,
  Bodnarchuk, Kovalenko, Rappe, and Lifshitz}}]{Isarov_5020_2017}
\bibinfo{author}{\bibfnamefont{M.}~\bibnamefont{Isarov}},
  \bibinfo{author}{\bibfnamefont{L.~Z.} \bibnamefont{Tan}},
  \bibinfo{author}{\bibfnamefont{M.~I.} \bibnamefont{Bodnarchuk}},
  \bibinfo{author}{\bibfnamefont{M.~V.} \bibnamefont{Kovalenko}},
  \bibinfo{author}{\bibfnamefont{A.~M.} \bibnamefont{Rappe}}, \bibnamefont{and}
  \bibinfo{author}{\bibfnamefont{E.}~\bibnamefont{Lifshitz}},
  \bibinfo{journal}{Nano Lett.} \textbf{\bibinfo{volume}{17}},
  \bibinfo{pages}{5020} (\bibinfo{year}{2017}).

\bibitem[{\citenamefont{Yang and Kelly}(2017)}]{Yang_92_2017}
\bibinfo{author}{\bibfnamefont{J.}~\bibnamefont{Yang}} \bibnamefont{and}
  \bibinfo{author}{\bibfnamefont{T.~L.} \bibnamefont{Kelly}},
  \bibinfo{journal}{Inorg. Chem.} \textbf{\bibinfo{volume}{56}},
  \bibinfo{pages}{92} (\bibinfo{year}{2017}).

\bibitem[{\citenamefont{Yaffe et~al.}(2017)\citenamefont{Yaffe, Guo, Tan,
  Egger, Hull, Stoumpos, Zheng, Heinz, Kronik, Kanatzidis
  et~al.}}]{Yaffe_136001_2017}
\bibinfo{author}{\bibfnamefont{O.}~\bibnamefont{Yaffe}},
  \bibinfo{author}{\bibfnamefont{Y.}~\bibnamefont{Guo}},
  \bibinfo{author}{\bibfnamefont{L.~Z.} \bibnamefont{Tan}},
  \bibinfo{author}{\bibfnamefont{D.~A.} \bibnamefont{Egger}},
  \bibinfo{author}{\bibfnamefont{T.}~\bibnamefont{Hull}},
  \bibinfo{author}{\bibfnamefont{C.~C.} \bibnamefont{Stoumpos}},
  \bibinfo{author}{\bibfnamefont{F.}~\bibnamefont{Zheng}},
  \bibinfo{author}{\bibfnamefont{T.~F.} \bibnamefont{Heinz}},
  \bibinfo{author}{\bibfnamefont{L.}~\bibnamefont{Kronik}},
  \bibinfo{author}{\bibfnamefont{M.~G.} \bibnamefont{Kanatzidis}},
  \bibnamefont{et~al.}, \bibinfo{journal}{Phys. Rev. Lett.}
  \textbf{\bibinfo{volume}{118}}, \bibinfo{pages}{136001}
  (\bibinfo{year}{2017}).

\bibitem[{\citenamefont{Sher et~al.}(1987)\citenamefont{Sher, van Schilfgaarde,
  Chen, and Chen}}]{Sher_4279_1987}
\bibinfo{author}{\bibfnamefont{A.}~\bibnamefont{Sher}},
  \bibinfo{author}{\bibfnamefont{M.}~\bibnamefont{van Schilfgaarde}},
  \bibinfo{author}{\bibfnamefont{A.}~\bibnamefont{Chen}}, \bibnamefont{and}
  \bibinfo{author}{\bibfnamefont{W.}~\bibnamefont{Chen}},
  \bibinfo{journal}{Phys. Rev. B} \textbf{\bibinfo{volume}{36}},
  \bibinfo{pages}{4279} (\bibinfo{year}{1987}).

\bibitem[{\citenamefont{Teles et~al.}(2000)\citenamefont{Teles, Furthm\"uller,
  Scolfaro, Leite, and Bechstedt}}]{Teles_2475_2000}
\bibinfo{author}{\bibfnamefont{L.~K.} \bibnamefont{Teles}},
  \bibinfo{author}{\bibfnamefont{J.}~\bibnamefont{Furthm\"uller}},
  \bibinfo{author}{\bibfnamefont{L.~M.~R.} \bibnamefont{Scolfaro}},
  \bibinfo{author}{\bibfnamefont{J.~R.} \bibnamefont{Leite}}, \bibnamefont{and}
  \bibinfo{author}{\bibfnamefont{F.}~\bibnamefont{Bechstedt}},
  \bibinfo{journal}{Phys. Rev. B} \textbf{\bibinfo{volume}{62}},
  \bibinfo{pages}{2475} (\bibinfo{year}{2000}).

\bibitem[{\citenamefont{Schleife et~al.}(2010)\citenamefont{Schleife,
  Eisenacher, {R\"odl}, Fuchs, {Furthm\"uller}, and
  Bechstedt}}]{Schleife_245210_2010}
\bibinfo{author}{\bibfnamefont{A.}~\bibnamefont{Schleife}},
  \bibinfo{author}{\bibfnamefont{M.}~\bibnamefont{Eisenacher}},
  \bibinfo{author}{\bibfnamefont{C.}~\bibnamefont{{R\"odl}}},
  \bibinfo{author}{\bibfnamefont{F.}~\bibnamefont{Fuchs}},
  \bibinfo{author}{\bibfnamefont{J.}~\bibnamefont{{Furthm\"uller}}},
  \bibnamefont{and}
  \bibinfo{author}{\bibfnamefont{F.}~\bibnamefont{Bechstedt}},
  \bibinfo{journal}{Phys. Rev. B} \textbf{\bibinfo{volume}{81}},
  \bibinfo{pages}{245210} (\bibinfo{year}{2010}).

\bibitem[{\citenamefont{Guilhon et~al.}(2015)\citenamefont{Guilhon, Teles,
  Marques, Pela, and Bechstedt}}]{Ghilhon_075435_2015}
\bibinfo{author}{\bibfnamefont{I.}~\bibnamefont{Guilhon}},
  \bibinfo{author}{\bibfnamefont{L.~K.} \bibnamefont{Teles}},
  \bibinfo{author}{\bibfnamefont{M.}~\bibnamefont{Marques}},
  \bibinfo{author}{\bibfnamefont{R.~R.} \bibnamefont{Pela}}, \bibnamefont{and}
  \bibinfo{author}{\bibfnamefont{F.}~\bibnamefont{Bechstedt}},
  \bibinfo{journal}{Phys. Rev. B} \textbf{\bibinfo{volume}{92}},
  \bibinfo{pages}{075435} (\bibinfo{year}{2015}).

\bibitem[{\citenamefont{Freitas et~al.}(2016)\citenamefont{Freitas,
  {Furthm\"uller}, Bechstedt, Marques, and Teles}}]{Freitas_092101_2016}
\bibinfo{author}{\bibfnamefont{F.~L.} \bibnamefont{Freitas}},
  \bibinfo{author}{\bibfnamefont{J.}~\bibnamefont{{Furthm\"uller}}},
  \bibinfo{author}{\bibfnamefont{F.}~\bibnamefont{Bechstedt}},
  \bibinfo{author}{\bibfnamefont{M.}~\bibnamefont{Marques}}, \bibnamefont{and}
  \bibinfo{author}{\bibfnamefont{L.~K.} \bibnamefont{Teles}},
  \bibinfo{journal}{Appl. Phys. Lett.} \textbf{\bibinfo{volume}{108}},
  \bibinfo{pages}{092101} (\bibinfo{year}{2016}).

\bibitem[{\citenamefont{Brivio et~al.}(2016)\citenamefont{Brivio, Caetano, and
  Walsh}}]{Brivio_1083_2016}
\bibinfo{author}{\bibfnamefont{F.}~\bibnamefont{Brivio}},
  \bibinfo{author}{\bibfnamefont{C.}~\bibnamefont{Caetano}}, \bibnamefont{and}
  \bibinfo{author}{\bibfnamefont{A.}~\bibnamefont{Walsh}}, \bibinfo{journal}{J.
  Phys. Chem. Lett.} \textbf{\bibinfo{volume}{7}}, \bibinfo{pages}{1083}
  (\bibinfo{year}{2016}).

\bibitem[{\citenamefont{Chen and Sher}(1995)}]{Chen_1995}
\bibinfo{author}{\bibfnamefont{A.~B.} \bibnamefont{Chen}} \bibnamefont{and}
  \bibinfo{author}{\bibfnamefont{A.}~\bibnamefont{Sher}},
  \emph{\bibinfo{title}{Semiconductor Alloys}} (\bibinfo{publisher}{Plenum, New
  York}, \bibinfo{year}{1995}), ISBN \bibinfo{isbn}{978-1-4613-0317-6}.

\bibitem[{\citenamefont{Leguy et~al.}(2015)\citenamefont{Leguy, Frost, McMahon,
  Sakai, Kockelmann, Law, Li, Foglia, Walsh, O$^\prime$Regan
  et~al.}}]{Leguy_1_2015}
\bibinfo{author}{\bibfnamefont{A.~M.~A.} \bibnamefont{Leguy}},
  \bibinfo{author}{\bibfnamefont{J.~M.} \bibnamefont{Frost}},
  \bibinfo{author}{\bibfnamefont{A.~P.} \bibnamefont{McMahon}},
  \bibinfo{author}{\bibfnamefont{V.~G.} \bibnamefont{Sakai}},
  \bibinfo{author}{\bibfnamefont{W.}~\bibnamefont{Kockelmann}},
  \bibinfo{author}{\bibfnamefont{C.}~\bibnamefont{Law}},
  \bibinfo{author}{\bibfnamefont{X.}~\bibnamefont{Li}},
  \bibinfo{author}{\bibfnamefont{F.}~\bibnamefont{Foglia}},
  \bibinfo{author}{\bibfnamefont{A.}~\bibnamefont{Walsh}},
  \bibinfo{author}{\bibfnamefont{B.~C.} \bibnamefont{O$^\prime$Regan}},
  \bibnamefont{et~al.}, \bibinfo{journal}{Nat. Commun.}
  \textbf{\bibinfo{volume}{6}}, \bibinfo{pages}{1} (\bibinfo{year}{2015}).

\bibitem[{\citenamefont{Chen et~al.}(2015)\citenamefont{Chen, Marco, Yang,
  Song, Chen, Zhao, Hong, Zhou, and Yang}}]{Chen_355_2015}
\bibinfo{author}{\bibfnamefont{Q.}~\bibnamefont{Chen}},
  \bibinfo{author}{\bibfnamefont{N.~D.} \bibnamefont{Marco}},
  \bibinfo{author}{\bibfnamefont{Y.~M.} \bibnamefont{Yang}},
  \bibinfo{author}{\bibfnamefont{T.}~\bibnamefont{Song}},
  \bibinfo{author}{\bibfnamefont{C.-C.} \bibnamefont{Chen}},
  \bibinfo{author}{\bibfnamefont{H.}~\bibnamefont{Zhao}},
  \bibinfo{author}{\bibfnamefont{Z.}~\bibnamefont{Hong}},
  \bibinfo{author}{\bibfnamefont{H.}~\bibnamefont{Zhou}}, \bibnamefont{and}
  \bibinfo{author}{\bibfnamefont{Y.}~\bibnamefont{Yang}},
  \bibinfo{journal}{Nano Today} \textbf{\bibinfo{volume}{10}},
  \bibinfo{pages}{355} (\bibinfo{year}{2015}), ISSN \bibinfo{issn}{1748-0132}.

\bibitem[{\citenamefont{Franssen et~al.}(2017)\citenamefont{Franssen, van Es,
  Dervi\c{s}o\u{g}lu, de~Wijs, and Kentgens}}]{Franssen_61_2017}
\bibinfo{author}{\bibfnamefont{W.~M.~J.} \bibnamefont{Franssen}},
  \bibinfo{author}{\bibfnamefont{S.~G.~D.} \bibnamefont{van Es}},
  \bibinfo{author}{\bibfnamefont{R.}~\bibnamefont{Dervi\c{s}o\u{g}lu}},
  \bibinfo{author}{\bibfnamefont{G.~A.} \bibnamefont{de~Wijs}},
  \bibnamefont{and} \bibinfo{author}{\bibfnamefont{A.~P.~M.}
  \bibnamefont{Kentgens}}, \bibinfo{journal}{J. Phys. Chem. Lett.}
  \textbf{\bibinfo{volume}{8}}, \bibinfo{pages}{61} (\bibinfo{year}{2017}).

\bibitem[{\citenamefont{Wang et~al.}(2018)\citenamefont{Wang, Lin, Zhu, Zheng,
  Wang, Li, and Zhu}}]{Wang_2772_2018}
\bibinfo{author}{\bibfnamefont{Y.}~\bibnamefont{Wang}},
  \bibinfo{author}{\bibfnamefont{R.}~\bibnamefont{Lin}},
  \bibinfo{author}{\bibfnamefont{P.}~\bibnamefont{Zhu}},
  \bibinfo{author}{\bibfnamefont{Q.}~\bibnamefont{Zheng}},
  \bibinfo{author}{\bibfnamefont{Q.}~\bibnamefont{Wang}},
  \bibinfo{author}{\bibfnamefont{D.}~\bibnamefont{Li}}, \bibnamefont{and}
  \bibinfo{author}{\bibfnamefont{J.}~\bibnamefont{Zhu}}, \bibinfo{journal}{Nano
  Lett.} \textbf{\bibinfo{volume}{18}}, \bibinfo{pages}{2772}
  (\bibinfo{year}{2018}).

\bibitem[{\citenamefont{{Dunlap-Shohl} et~al.}(0)\citenamefont{{Dunlap-Shohl},
  Zhou, Padture, and Mitzi}}]{DunlapShohl_XXX_2018}
\bibinfo{author}{\bibfnamefont{W.~A.} \bibnamefont{{Dunlap-Shohl}}},
  \bibinfo{author}{\bibfnamefont{Y.}~\bibnamefont{Zhou}},
  \bibinfo{author}{\bibfnamefont{N.~P.} \bibnamefont{Padture}},
  \bibnamefont{and} \bibinfo{author}{\bibfnamefont{D.~B.} \bibnamefont{Mitzi}},
  \bibinfo{journal}{Chem. Rev.} \textbf{\bibinfo{volume}{0}},
  \bibinfo{pages}{null} (\bibinfo{year}{0}).

\bibitem[{\citenamefont{Even et~al.}(2012)\citenamefont{Even, Pedesseau,
  Dupertuis, Jancu, and Katan}}]{Even_205301_2012}
\bibinfo{author}{\bibfnamefont{J.}~\bibnamefont{Even}},
  \bibinfo{author}{\bibfnamefont{L.}~\bibnamefont{Pedesseau}},
  \bibinfo{author}{\bibfnamefont{M.}~\bibnamefont{Dupertuis}},
  \bibinfo{author}{\bibfnamefont{J.}~\bibnamefont{Jancu}}, \bibnamefont{and}
  \bibinfo{author}{\bibfnamefont{C.}~\bibnamefont{Katan}},
  \bibinfo{journal}{Phys. Rev. B} \textbf{\bibinfo{volume}{86}},
  \bibinfo{pages}{205301} (\bibinfo{year}{2012}).

\bibitem[{\citenamefont{Even et~al.}(2013)\citenamefont{Even, Pedesseau, Jancu,
  and Katan}}]{Even_2999_2013}
\bibinfo{author}{\bibfnamefont{J.}~\bibnamefont{Even}},
  \bibinfo{author}{\bibfnamefont{L.}~\bibnamefont{Pedesseau}},
  \bibinfo{author}{\bibfnamefont{J.-M.} \bibnamefont{Jancu}}, \bibnamefont{and}
  \bibinfo{author}{\bibfnamefont{C.}~\bibnamefont{Katan}}, \bibinfo{journal}{J.
  Phys. Chem. Lett.} \textbf{\bibinfo{volume}{4}}, \bibinfo{pages}{2999}
  (\bibinfo{year}{2013}).

\bibitem[{\citenamefont{Even et~al.}(2015)\citenamefont{Even, Pedesseau, Katan,
  Kepenekian, Lauret, Sapori, and Deleporte}}]{Even_10161_2015}
\bibinfo{author}{\bibfnamefont{J.}~\bibnamefont{Even}},
  \bibinfo{author}{\bibfnamefont{L.}~\bibnamefont{Pedesseau}},
  \bibinfo{author}{\bibfnamefont{C.}~\bibnamefont{Katan}},
  \bibinfo{author}{\bibfnamefont{M.}~\bibnamefont{Kepenekian}},
  \bibinfo{author}{\bibfnamefont{J.}~\bibnamefont{Lauret}},
  \bibinfo{author}{\bibfnamefont{D.}~\bibnamefont{Sapori}}, \bibnamefont{and}
  \bibinfo{author}{\bibfnamefont{E.}~\bibnamefont{Deleporte}},
  \bibinfo{journal}{J. Phys. Chem. C} \textbf{\bibinfo{volume}{119}},
  \bibinfo{pages}{10161} (\bibinfo{year}{2015}).

\bibitem[{\citenamefont{Kittel}(2004)}]{Kittel_2004}
\bibinfo{author}{\bibfnamefont{C.}~\bibnamefont{Kittel}},
  \emph{\bibinfo{title}{Introduction to Solid State Physics, 8th Ed.}}
  (\bibinfo{publisher}{John Wiley \& Sons, Inc., New York},
  \bibinfo{year}{2004}), ISBN \bibinfo{isbn}{047141526X},
  \urlprefix\url{http://www.ebook.de/de/product/4290142/charles_kittel_introduction_to_solid_state_physics.html}.

\bibitem[{\citenamefont{Weller et~al.}(2015)\citenamefont{Weller, Weber, Henry,
  Di~Pumpo, and Hansen}}]{Weller_4180_2015}
\bibinfo{author}{\bibfnamefont{M.~T.} \bibnamefont{Weller}},
  \bibinfo{author}{\bibfnamefont{O.~J.} \bibnamefont{Weber}},
  \bibinfo{author}{\bibfnamefont{P.~F.} \bibnamefont{Henry}},
  \bibinfo{author}{\bibfnamefont{A.~M.} \bibnamefont{Di~Pumpo}},
  \bibnamefont{and} \bibinfo{author}{\bibfnamefont{T.~C.}
  \bibnamefont{Hansen}}, \bibinfo{journal}{Chem. Commun.}
  \textbf{\bibinfo{volume}{51}}, \bibinfo{pages}{4180} (\bibinfo{year}{2015}).

\bibitem[{\citenamefont{Krishnamoorthy
  et~al.}(2015)\citenamefont{Krishnamoorthy, Ding, Yan, Leong, Baikie, Zhang,
  Sherburne, Li, Asta, Mathews et~al.}}]{Krishnamoorthy_23829_2015}
\bibinfo{author}{\bibfnamefont{T.}~\bibnamefont{Krishnamoorthy}},
  \bibinfo{author}{\bibfnamefont{H.}~\bibnamefont{Ding}},
  \bibinfo{author}{\bibfnamefont{C.}~\bibnamefont{Yan}},
  \bibinfo{author}{\bibfnamefont{W.~L.} \bibnamefont{Leong}},
  \bibinfo{author}{\bibfnamefont{T.}~\bibnamefont{Baikie}},
  \bibinfo{author}{\bibfnamefont{Z.}~\bibnamefont{Zhang}},
  \bibinfo{author}{\bibfnamefont{M.}~\bibnamefont{Sherburne}},
  \bibinfo{author}{\bibfnamefont{S.}~\bibnamefont{Li}},
  \bibinfo{author}{\bibfnamefont{M.}~\bibnamefont{Asta}},
  \bibinfo{author}{\bibfnamefont{N.}~\bibnamefont{Mathews}},
  \bibnamefont{et~al.}, \bibinfo{journal}{J. Mater. Chem. A}
  \textbf{\bibinfo{volume}{3}}, \bibinfo{pages}{22829} (\bibinfo{year}{2015}).

\bibitem[{\citenamefont{Tao et~al.}(2017)\citenamefont{Tao, Cao, and
  Bobbert}}]{Tao_1_2017}
\bibinfo{author}{\bibfnamefont{S.~X.} \bibnamefont{Tao}},
  \bibinfo{author}{\bibfnamefont{X.}~\bibnamefont{Cao}}, \bibnamefont{and}
  \bibinfo{author}{\bibfnamefont{P.~A.} \bibnamefont{Bobbert}},
  \bibinfo{journal}{Sci. Rep.} \textbf{\bibinfo{volume}{7}}, \bibinfo{pages}{1}
  (\bibinfo{year}{2017}).

\bibitem[{\citenamefont{Jiang et~al.}(2017)\citenamefont{Jiang, Wu, Sun, Li,
  Li, Lu, Zou, and Deng}}]{Jiang_24359_2017}
\bibinfo{author}{\bibfnamefont{L.}~\bibnamefont{Jiang}},
  \bibinfo{author}{\bibfnamefont{T.}~\bibnamefont{Wu}},
  \bibinfo{author}{\bibfnamefont{L.}~\bibnamefont{Sun}},
  \bibinfo{author}{\bibfnamefont{Y.-J.} \bibnamefont{Li}},
  \bibinfo{author}{\bibfnamefont{A.}~\bibnamefont{Li}},
  \bibinfo{author}{\bibfnamefont{R.}~\bibnamefont{Lu}},
  \bibinfo{author}{\bibfnamefont{K.}~\bibnamefont{Zou}}, \bibnamefont{and}
  \bibinfo{author}{\bibfnamefont{W.-Q.} \bibnamefont{Deng}},
  \bibinfo{journal}{J. Phys. Chem. C} \textbf{\bibinfo{volume}{121}},
  \bibinfo{pages}{24359} (\bibinfo{year}{2017}).

\bibitem[{\citenamefont{Denton and Ashcroft}(1991)}]{Denton_3161_1991}
\bibinfo{author}{\bibfnamefont{A.~R.} \bibnamefont{Denton}} \bibnamefont{and}
  \bibinfo{author}{\bibfnamefont{N.~W.} \bibnamefont{Ashcroft}},
  \bibinfo{journal}{Phys. Rev. A} \textbf{\bibinfo{volume}{43}},
  \bibinfo{pages}{3161} (\bibinfo{year}{1991}),
  \urlprefix\url{http://dx.doi.org/10.1103/PhysRevA.43.3161}.

\bibitem[{\citenamefont{Carignano et~al.}(2017)\citenamefont{Carignano,
  Aravindh, Roqan, Even, and Katan}}]{Carignano_20729_2017}
\bibinfo{author}{\bibfnamefont{M.~A.} \bibnamefont{Carignano}},
  \bibinfo{author}{\bibfnamefont{S.~A.} \bibnamefont{Aravindh}},
  \bibinfo{author}{\bibfnamefont{I.~S.} \bibnamefont{Roqan}},
  \bibinfo{author}{\bibfnamefont{J.}~\bibnamefont{Even}}, \bibnamefont{and}
  \bibinfo{author}{\bibfnamefont{C.}~\bibnamefont{Katan}}, \bibinfo{journal}{J.
  Phys. Chem. C} \textbf{\bibinfo{volume}{121}}, \bibinfo{pages}{20729}
  (\bibinfo{year}{2017}).

\bibitem[{\citenamefont{Hohenberg and Kohn}(1964)}]{Hohenberg_B864_1964}
\bibinfo{author}{\bibfnamefont{P.}~\bibnamefont{Hohenberg}} \bibnamefont{and}
  \bibinfo{author}{\bibfnamefont{W.}~\bibnamefont{Kohn}},
  \bibinfo{journal}{Phys. Rev.} \textbf{\bibinfo{volume}{136}},
  \bibinfo{pages}{B864} (\bibinfo{year}{1964}),
  \urlprefix\url{http://dx.doi.org/10.1103/PhysRev.136.B864}.

\bibitem[{\citenamefont{Kohn and Sham}(1965)}]{Kohn_A1133_1965}
\bibinfo{author}{\bibfnamefont{W.}~\bibnamefont{Kohn}} \bibnamefont{and}
  \bibinfo{author}{\bibfnamefont{L.~J.} \bibnamefont{Sham}},
  \bibinfo{journal}{Phys. Rev.} \textbf{\bibinfo{volume}{140}},
  \bibinfo{pages}{A1133} (\bibinfo{year}{1965}),
  \urlprefix\url{http://dx.doi.org/10.1103/PhysRev.140.A1133}.

\bibitem[{\citenamefont{Perdew et~al.}(1996)\citenamefont{Perdew, Burke, and
  Ernzerhof}}]{Perdew_3865_1996}
\bibinfo{author}{\bibfnamefont{J.~P.} \bibnamefont{Perdew}},
  \bibinfo{author}{\bibfnamefont{K.}~\bibnamefont{Burke}}, \bibnamefont{and}
  \bibinfo{author}{\bibfnamefont{M.}~\bibnamefont{Ernzerhof}},
  \bibinfo{journal}{Phys. Rev. Lett.} \textbf{\bibinfo{volume}{77}},
  \bibinfo{pages}{3865} (\bibinfo{year}{1996}),
  \urlprefix\url{http://dx.doi.org/10.1103/PhysRevLett.77.3865}.

\bibitem[{\citenamefont{Bl{\"o}chl}(1994)}]{Blochl_17953_1994}
\bibinfo{author}{\bibfnamefont{P.~E.} \bibnamefont{Bl{\"o}chl}},
  \bibinfo{journal}{Phys. Rev. B} \textbf{\bibinfo{volume}{50}},
  \bibinfo{pages}{17953} (\bibinfo{year}{1994}),
  \urlprefix\url{http://dx.doi.org/10.1103/PhysRevB.50.17953}.

\bibitem[{\citenamefont{Kresse and Joubert}(1999)}]{Kresse_1758_1999}
\bibinfo{author}{\bibfnamefont{G.}~\bibnamefont{Kresse}} \bibnamefont{and}
  \bibinfo{author}{\bibfnamefont{D.}~\bibnamefont{Joubert}},
  \bibinfo{journal}{Phys. Rev. B} \textbf{\bibinfo{volume}{59}},
  \bibinfo{pages}{1758} (\bibinfo{year}{1999}),
  \urlprefix\url{http://dx.doi.org/10.1103/PhysRevB.59.1758}.

\bibitem[{\citenamefont{Kresse and Hafner}(1993)}]{Kresse_13115_1993}
\bibinfo{author}{\bibfnamefont{G.}~\bibnamefont{Kresse}} \bibnamefont{and}
  \bibinfo{author}{\bibfnamefont{J.}~\bibnamefont{Hafner}},
  \bibinfo{journal}{Phys. Rev. B} \textbf{\bibinfo{volume}{48}},
  \bibinfo{pages}{13115} (\bibinfo{year}{1993}),
  \urlprefix\url{http://dx.doi.org/10.1103/PhysRevB.48.13115}.

\bibitem[{\citenamefont{Kresse and Furthm{\"u}ller}(1996)}]{Kresse_11169_1996}
\bibinfo{author}{\bibfnamefont{G.}~\bibnamefont{Kresse}} \bibnamefont{and}
  \bibinfo{author}{\bibfnamefont{J.}~\bibnamefont{Furthm{\"u}ller}},
  \bibinfo{journal}{Phys. Rev. B} \textbf{\bibinfo{volume}{54}},
  \bibinfo{pages}{11169} (\bibinfo{year}{1996}),
  \urlprefix\url{http://dx.doi.org/10.1103/PhysRevB.54.11169}.

\end{thebibliography}

\section*{Acknowledgements}
We thank the Brazilian funding agency Coordination for Improvement of Higher Level 
Education (CAPES), PVE Grants No. 88881.068355/2014-01 and No. 88887.145962/2017-00,
the National Council for Scientific and Technological Development (CNPq), Grants 
No. 308742/2016-8, and the S\~ao Paulo Research Foundation (FAPESP) for the financial 
supports.
We also thank the Scientific Computation National Laboratory for provide Santos 
Dumont Supercomputer resources to perform the electronic structure calculations.



\end{document}